\input amstex
\documentstyle{amsppt}

\mag=\magstep1
\vsize=21.6truecm
\hsize=16truecm
\NoBlackBoxes
\leftheadtext{Yunbo Zeng}
\rightheadtext{A method to introduce additional separated variables }

\topmatter
 \title A new method to introduce additional separated
 variables for high-order binary constrained flows
\endtitle
\author
Yunbo Zeng \footnote { E-mail:yzeng\@tsinghua.edu.cn}
\endauthor
\affil
Department of Mathematical Sciences,
Tsinghua University\\ Beijing 100084, China\\
\endaffil
\endtopmatter
\def\p{{\partial}}
\def\la{{\lambda}}
\def\La{{\Lambda}}

\TagsOnRight
{\bf Abstract.} Degrees of freedom for high-order binary constrained
flows
of
soliton equations admitting
$2\times 2$ Lax matrices are $2N+k_0$. It is known that $N+k_0$
pairs of
canonical separated variables for their separation
 of variables can be introduced directly via
 their Lax matrices.
 In present paper we propose a new method to
 introduce the additional $N$ pairs of canonical separated variables
 and $N$ additional separated equations.
 The Jacobi inversion problems for high-order
 binary constrained flows  and for soliton equations are also
established.
This new method can be applied to all high-order binary constrained
flows admitting $2\times 2$ Lax matrices.
\par
\  \par
\  \par
\ \par
Classification numbers: 02.90.+p

\ \par

\newpage

\subhead {1. Introduction}\endsubhead\par

For a finite-dimensional integrable Hamiltonian systems (FDIHS),
 let $m$ denote the number of degrees of freedom, and $P_i, i=1,...,m,$
be
functionally independent integrals of motion in involution, the
separation
of
variables means to
 construct $m$ pairs of canonical separated variables $v_k, u_k,
k=1,...,m$,
 [1,2,3]
$$\{u_k, u_l\}=\{v_k, v_l\}=0,\qquad \{v_k, u_l\}=\delta_{kl},\qquad
k,l=1,...,m, \tag 1.1$$
and $m$ functions $f_k$ such that
$$f_k(u_k, v_k, P_1,...,P_{m})=0, \qquad k=1,...,m,\tag 1.2$$
which  are called separated equations. The equations (1.2) give rise to
an explicit factorization of the Liouville tori.
For the FDIHSs with the Lax matrices admitting the $r$-matrices of the
$XXX, XXZ$ and $XYZ$ type, there is a general approach to their
separation of variables
[1-6]. The corresponding separated equations enable us to express
the generating function of canonical transformation in completely
separated form as an abelian integral on the associated invariant
spectral curve. The resulted linearizing map is essentially the Abel
map to the Jacobi variety of the spectral curve, thus providing a link
 with the algebro-geometric
linearization methods given by [7-9].\par
The separation of variables for a FDIHS requires
 that the number of canonical separated variables $u_k$ should be
 equal to the number $m$ of degrees of freedom. In some cases, the
 number of $u_k$ resulted by the normal method may be less than $m$
 and one needs to introduce some additional canonical separated
variables.
 So far very few models in these cases have been studied. These cases
remain
 to be a challenging problem [3].\par
 The separation of variables for constrained flows of soliton equations
 has been studied (see, for example, [4,10-14]).
In recent years binary constrained flows of soliton hierarchies have
attracted  attention (see, for example, [15-22]). However the separation

of
variables for binary constrained flows has not been studied.
The degree of freedom
for high-order binary constrained flows admitting $2\times 2$ Lax
matrices
$M=\left( \matrix A(\la)&B(\la)\\C(\la)&-A(\la)\endmatrix\right)$
is a natural number $2N+k_0$. Via the Lax matrix $M$,
$N+k_0$ pairs of canonical separated variables $u_1,...,u_{N+k_0}$ can
be
introduced
 by the set of zeros of $B(\la)$ and $v_k=A(u_k)$, and
$N+k_0$ separated equations can be found from the
generation function of
intrgrals of motions.
In previous papers [23,24] we pesented a method with two different ways
for determining  additional $N$ pairs of canonical separated variables
and
additional $N$ separated equations for first binary constrained flows
with
$2N$ degree of freedom.
The main idea in [23,24] is to
construct two functions $\widetilde B(\la)$ and $\widetilde A(\la)$
and define $u_{N+1},...,u_{2N}$ by the set of zeros of $\widetilde
B(\la)$
and $v_{N+k}=\widetilde A(u_{N+k})$. The ways for
constructing $\widetilde B(\la)$ and $\widetilde A(\la)$  in [23] and
[24]
are some different.
In present paper we propose a completely different method from that in
[23,24] to introduce the additional
$N$ separated variables and $N$ separated equations for high-order
binary constrained flows with $2N+k_0$ degree of freedom.
It is observed
that
the introduction of $v_k$ has some link with integrals of motion and
should lead to the separated equations. We find that there are $N$
integrals
of motion $P_{N+k_0+1},...,P_{2N+k_0}$
among the $2N+k_0$ integrals of motion for the high-order binary
constrained flows
which commute with $A(\la)$ and
$B(\la)$. This observation and property stimulate us
to directly use the additional integrals of motion to define both
the $N$ pairs of additional separated
variables and $N$ separated equations by $v_{N+k_0+j}=P_{N+k_0+j},
j=1,...,N$.
Then we can find
the conjugated variables $u_{N+k_0+j}, 1,...,N,$
commuting with $A(\la)$
and $B(\la)$.
In contrast to the method in [23,24], this method
is easer to be applied
to the
high-order binary constrained flows.\par
We will also present
the separation of variables of soliton equations.
The first step is
to factorize $(1+1)-$dimensional soliton equations into
two commuting $x-$ and $t-$FDIHSs via high-order binary constrained
flows,
namely the $x-$ and $t-$dependences of the soliton equations are
separated by the $x-$ and $t-$FDIHSs obtained from the $x$- and
$t$-binary constrained flows. The second step is to produce separation
of variables for the $x-$ and $t-$FDIHDs. Finally, combining the
factorization of soliton equations
with the Jacobi inversion problems for $x-$ and $t-$FDIHSs enables
us to establish the Jacobi inversion problems for soliton equations.
If the Jacobi inversion problem can be solved by the Jacobi inversion
technique [7], one can obtain the solution in terms
of the Riemann-theta
function for soliton equations.
We illustrate the method by KdV, AKNS and Kaup-Newell (KN) hierarchies.
The paper is organized as follows.\par
In section 2, we first recall the high-order binary constrained flows
and
 factorization of the KdV hierarchy. Then propose the mehtod for
 introducing the $N$ pairs of additional separated variables.
 We illustrate the method by both first binary constrained flow and
 second binary constrained flow. Finally we present the
 separation of variables for KdV hierarchy.
In section 3 and 4, the method is applied to the AKNS hierarchy and
KN hierarchy, respectively.
In fact this method can be applied to all high-order binary constrained
flows and other soliton hierarchies admitting $2\times 2$ Lax pairs.\par

\par
\subhead {2. Separation of variables for the KdV
equations}\endsubhead\par

We first recall the high-order binary constrained flows of the KdV
hierarchy.
\subhead {2.1 High-order binary constrained flows of the KdV hierarchy}
\endsubhead\par
Consider the Schr$\ddot{\text o}$dinger equation [25]
$$\phi_{xx}+(\la+u)\phi=0$$
which is equivalent to the following spectral problem
$$\phi_x=U(u,\la)\phi,\quad
 U(u, \lambda)
=\left( \matrix 0&1\\-\la-u&0\endmatrix\right),\quad
\phi=\binom {\phi_{1}}{\phi_{2}}.\tag 2.1$$
Take the time evolution law of $\phi$ as
$$\phi_{t_n}=V^{(n)}(u,\la)\phi,\tag 2.2$$
where
$$V^{(n)}(u, \la)=\sum_{i=0}^{n+1}
\left( \matrix a_i&b_i\\c_i&-a_i\endmatrix\right)\la^{n+1-i}
+\left( \matrix 0&0\\b_{n+2}&0\endmatrix\right),$$
$$a_{0}=b_{0}=0,\quad c_{0}=-1,\quad a_1=0,\quad b_{1}=1,\quad$$
$$ b_{k+1}=Lb_{k}=-\frac 12L^{k-1}u,, \quad
a_k=-\frac 12b_{k,x},\quad
c_k=-\frac 12b_{k,xx}-b_{k+1}-b_{k}u,\quad
 k=1,2,\cdots,$$
$$L=-\frac 14\p^2-u+\frac 12\p^{-1}u_x,\qquad \p=\p_x,
\quad \p^{-1}\p=\p\p^{-1}=1.\tag 2.3$$\par
The compatibility condition of (2.1) and (2.2) gives rise
to the n-th KdV equation which can be written as
an infinite-dimensional
Hamiltonian system [25]
$$u_{t_n}=-2b_{n+2,x}=\p L^nu=\p\frac {\delta H_{n}}{\delta u}, \tag
2.4$$
with the Hamiltonian
$H_{n}=\frac{4b_{n+3}}{2n+3}$ and $
\frac {\delta H_{n}}{\delta u}=-2b_{n+2}.$\par
 For $n=1$ we have
$$\phi_{t_1}=V^{(1)}(u,\la)\phi,\qquad
V^{(1)}=\left( \matrix \frac 14u_x&\la-\frac 12u\\-\la^2-\frac 12u\la
+\frac 14u_{xx}+\frac 12u^2&-\frac 14u_x\endmatrix\right),\tag 2.5$$
and the equation (2.4) for $n=1$ is the well-known KdV equation
$$u_{t_1}=-\frac 14(u_{xxx}+6uu_x). \tag 2.6$$\par
The adjoint spectral problem reads
$$\psi_x=-U^T(u,\la)\psi,\qquad
\psi=\binom {\psi_{1}}{\psi_{2}}.\tag 2.7$$
By means of the formula in [26], we have
$$\frac {\delta\la}{\delta u}
=Tr[\left(\matrix \phi_1\psi_1
& \phi_1\psi_2\\ \phi_2\psi_1&\phi_2\psi_2
\endmatrix\right)\frac {\p U(u,\la)}{\p u}]=-\psi_2\phi_1.$$\par
According to [15-22], the high-order binary $x$-constrained flows
of the
KdV hierarchy (2.4) consist
 of the equations obtained from
 the spectral problem  (2.1) and the adjoint spectral problem (2.7) for
$N$
distinct real numbers $\lambda_j$ and the restriction of the variational

derivatives for the conserved
quantities $H_{k_0}$ (for any fixed $k_0$) and $\lambda_{j}$:
$$ \Phi_{1,x}=\Phi_{2},\qquad\Phi_{2,x}=-\La\Phi_{1}-u\Phi_{1}
,\tag 2.8a$$
$$ \Psi_{1,x}=\La\Psi_{2}+u\Psi_{2},\qquad\Psi_{2,x}=-\Psi_{1},\tag
2.8b$$
$$\frac {\delta H_{k_0}}{\delta u}-
\sum_{j=1}^{N}\frac {\delta \lambda_{j}}{\delta u}
=-2b_{k_0+2}+<\Psi_2,\Phi_1>=0.\tag 2.8c$$
Hereafter we denote the inner product in $\text{\bf R}^N$ by
$<.,.>$
and
$$\Phi_i=(\phi_{i1},\cdots,\phi_{iN})^{T},\quad
\Psi_i=(\psi_{i1},\cdots,\psi_{iN})^{T},\quad i=1,2,
\quad  \Lambda=diag
(\lambda_1,\cdots,\lambda_N).$$\par
The binary $t_n$-constrained flows of the KdV hierarchy (2.4)
are defined by the replicas of
(2.2) and its adjoint system for $N$
distinct real number $\lambda_j$
$$ \binom {\phi_{1j}}{\phi_{2j}}_{t_n}
=V^{(n)}(u,\la_j)\binom {\phi_{1j}}{\phi_{2j}},
\quad\binom {\psi_{1j}}{\psi_{2j}}_{t_n}=-(V^{(n)}(u,\la_j))^T
\binom {\psi_{1j}}{\psi_{2j}},\quad j=1,...,N, \tag 2.9a$$
as well as the n-th KdV equation itself (2.4)
in the case  of the higher-order constraint for $k_0\ge 1$
$$u_{t_n}=-2b_{n+2,x}. \tag 2.9b$$\par
(1) For $k_0=0$, we have
$$b_{2}=-\frac 12u=\frac 12<\Psi_2,\Phi_1>,\quad i.e.,
\quad u=-<\Psi_2,\Phi_1>.\tag 2.10$$
By substituting (2.10), (2.8a) and (2.8b) becomes a
 finite-dimensional Hamiltonian system (FDHS) [18]
$$ \Phi_{1x}=\frac {\p F_1}{\p \Psi_1},\quad \Phi_{2x}
=\frac {\p F_1}{\p \Psi_2},\quad
 \Psi_{1x}=-\frac {\p F_1}{\p \Phi_1},\quad
 \Psi_{2x}=-\frac {\p F_1}{\p \Phi_2},\tag 2.11$$
$$F_1=<\Psi_1, \Phi_2>-<\La\Psi_2, \Phi_1>
+\frac 12<\Psi_2, \Phi_1>^2.$$\par
Under the constraint (2.10) and the $x$-FDHS (2.11), the binary
$t_1$-constrained flow obtained from  (2.9a) with $V^{(1)}$ given
by (2.5) can also be written as a $t_1$-FDHS
$$ \Phi_{1,t_1}=\frac {\p F_2}{\p \Psi_1},\quad \Phi_{2,t_1}
=\frac {\p F_2}{\p \Psi_2},\quad
 \Psi_{1,t_1}=-\frac {\p F_2}{\p \Phi_1},\quad
 \Psi_{2,t_1}=-\frac {\p F_2}{\p \Phi_2},\tag 2.12$$
$$F_2=-<\La^2\Psi_2, \Phi_1>+<\La\Psi_1, \Phi_2>
+\frac 12<\Psi_2, \Phi_1><\La \Psi_2, \Phi_1>$$
$$+\frac 12<\Psi_2, \Phi_1><\Psi_1, \Phi_2>
+\frac 18(<\Psi_2, \Phi_2>-<\Psi_1, \Phi_1>)^2.$$\par
The Lax representation for the $x$-constrained flow (2.8) and the
$t_n$-constrained flow (2.9) can be deduced from the adjoint
representation
of (2.1) and (2.2) by using the method in [27,28]
$$M_x=[\widetilde U, M],\qquad M_{t_n}=[\widetilde V^{(n)}, M],\tag
2.13$$
where $\widetilde U$ and $\widetilde V^{(n)}$ are obtained from $U$
and $V^{(n)}$ under the system (2.8), and the Lax matrix $M$ is
of the form
$$M=\left( \matrix A(\la)&B(\la)\\C(\la)&-A(\la)\endmatrix\right).$$\par

The Lax matrix $M$ for $x$-FDHS (2.11) and $t_1$-FDHS (2.12) is given by

$$A(\la)=\frac{1}{4}\sum_{j=1}^{N}\frac{\psi_{1j}\phi_{1j}
-\psi_{2j}\phi_{2j}}{\la-\la_{j}},\qquad
B(\la)=1+\frac 12\sum_{j=1}^{N}\frac{\psi_{2j}\phi_{1j}}{\la-\la_{j}},$$

$$C(\la)=-\la+\frac 12<\Psi_2, \Phi_1>
+\frac 12\sum_{j=1}^{N}\frac{\psi_{1j}\phi_{2j}}{\la-\la_{j}}.
\tag 2.14$$
The  generating function of  integrals of motion for (2.11) and (2.12)
yields
$$A^2(\la)+B(\la)C(\la)\equiv P(\la)=-\la+
\sum_{j=1}^{N}[\frac{P_{j}}{\la-\la_{j}}
+\frac{P^2_{N+j}}{(\la-\la_{j})^2}], \tag 2.15$$
where $P_1,...,P_{2N}$ are independent integrals of
motion for the FDHSs
(2.11) and (2.12)
$$P_j=\frac 12\psi_{1j}\phi_{2j}+(-\frac 12\la_j
+\frac 14<\Psi_2, \Phi_1>)\psi_{2j}\phi_{1j}$$
$$+\frac 18\sum_{k\neq j}\frac{1}{\la_j-\la_{k}}[(\psi_{1j}\phi_{1j}
-\psi_{2j}\phi_{2j})(\psi_{1k}\phi_{1k}
-\psi_{2k}\phi_{2k})+4\psi_{1j}\phi_{2j}\psi_{2k}\phi_{1k}],\tag 2.16a$$

$$P_{N+j}=\frac 14(\psi_{1j}\phi_{1j}+\psi_{2j}\phi_{2j}),\qquad
j=1,...,N.\tag 2.16b$$
It is easy to verify that
$$F_1=2\sum_{j=1}^{N}P_{j}, \qquad
F_2=2\sum_{j=1}^{N}(\la_jP_{j}+P^2_{N+j}). \tag 2.17$$\par
With respect to the standard Poisson bracket
$$\{f, g\}= \sum_{j=1}^{N}(\frac {\p f}{\p \psi_{1j}}
\frac {\p g}{\p \phi_{1j}}
+\frac {\p f}{\p \psi_{2j}} \frac {\p g}{\p \phi_{2j}}
-\frac {\p f}{\p \phi_{1j}} \frac {\p g}{\p \psi_{1j}}
-\frac {\p f}{\p \phi_{2j}} \frac {\p g}{\p \psi_{2j}}),\tag 2.18$$
by calculating the formulas like (2.31), it is easy to verify that
$$\{A^2(\la)+B(\la)C(\la), A^2(\mu)+B(\mu)C(\mu)\}=0,\tag 2.19$$
which implies that $P_1,..,P_{2N}$ are in involution, (2.11) and (2.12)
are
FDIHSs and
commute with each other.
The construction of (2.11) and (2.12) ensures that if
$(\Psi_1, \Psi_2, \Phi_1,\Phi_2)$ satisfies the FDIHSs (2.11)
and (2.12) simultaneously, then
$u$ defined by (2.10) solves the KdV equation (2.6).\par  Set
$$A^2(\la)+B(\la)C(\la)=
\la\sum_{k=0}^{\infty}\widetilde F_k\la^{-k}, \tag 2.20$$
where $\widetilde F_k, k=1,2,...,$ are also integrals of motion for
both the $x$-FDHSs (2.11) and the $t_n$-binary constrained flows (2.9).
Comparing (2.20) with (2.15), one gets
$$\widetilde F_0=-1,\quad \widetilde F_1=0, \quad \widetilde F_k
=\sum_{j=1}^{N}[\la_j^{k-2}P_j+(k-2)\la_j^{k-3}P_{N+j}^2],\quad
k=2,3,....
\tag 2.21$$
By employing the method in [28,29], the $t_n$-FDIHS obtained from
the $t_n$-binary constrained flow (2.9) is found to be of the form
$$ \Phi_{1,t_n}=\frac {\p F_{n+1}}{\p \Psi_1},\quad \Phi_{2,t_n}
=\frac {\p F_{n+1}}{\p \Psi_2},\quad
 \Psi_{1,t_n}=-\frac {\p F_{n+1}}{\p \Phi_1},\quad
 \Psi_{2,t_n}=-\frac {\p F_{n+1}}{\p \Phi_2},\tag 2.22a$$
$$F_{n+1}=\sum_{m=0}^{n}(\frac 12)^{m-1}\frac{\alpha_m}{m+1}
\sum_{l_1+...+l_{m+1}=n+2}\widetilde F_{l_1}...\widetilde F_{l_{m+1}},
\tag 2.22b$$
where $l_1\geq 1,...,l_{m+1}\geq 1, \alpha_0=1, \alpha_1=\frac 12,
\alpha_2=\frac 32,$ and [28,29]
$$\alpha_m=2\alpha_{m-1}+\sum_{l=1}^{m-2}\alpha_{l}\alpha_{m-l-1}
-\frac 12\sum_{l=1}^{m-1}\alpha_{l}\alpha_{m-l}, \quad m\geq 3.\tag
2.22c $$
The n-th KdV equation (2.4) is factorized by the $x$-FDIHS (2.11)
and the $t_n$-FDIHS (2.22).\par

(2) For $k_0=1$, one gets
$$b_{3}=\frac 18(u_{xx}+3u^2)=\frac 12<\Psi_2,\Phi_1>.\tag 2.23$$
By introducing $q=u, p=\frac 14 u_x,$ (2.8a), (2.8b) and (2.23) can be
written as a $x$-FDHS
$$ \Phi_{ix}=\frac {\p F_1}{\p \Psi_i},\quad \Psi_{ix}
=-\frac {\p F_1}{\p \Phi_i},\quad i=1,2, \quad
 q_{x}=\frac {\p F_1}{\p p},\quad
 p_{x}=-\frac {\p F_1}{\p q},\tag 2.24$$
$$F_1=-<\La\Psi_2, \Phi_1>+<\Psi_1, \Phi_2>
-q<\Psi_2, \Phi_1>+2p^2+\frac 14 q^3.$$\par
Under the constraint (2.23), $V^{(1)}$ becomes
$$\widetilde V^{(1)}=\left( \matrix  p&\la-\frac 12q\\
-\la^2-\frac 12q\la+<\Psi_2, \Phi_1>-
\frac 14q^2&-p\endmatrix\right).\tag 2.25$$
Under the constraint (2.23) and the $x$-FDHS (2.24),
the binary $t_1$-constrained flow consists of (2.9a) with $V^{(1)}$
replaced by $\widetilde V^{(1)}$ and  (2.9b) given by (2.6) can
also be written as a $t_1$-FDHS
$$ \Phi_{it_1}=\frac {\p F_2}{\p \Psi_i},\quad \Psi_{it_1}
=-\frac {\p F_2}{\p \Phi_i},\quad i=1,2, \quad
 q_{t_1}=\frac {\p F_2}{\p p},\quad
 p_{t_1}=-\frac {\p F_2}{\p q},\tag 2.26$$
$$F_2=-<\La^2\Psi_2, \Phi_1>+<\La\Psi_1, \Phi_2>
-\frac 12 q<\La\Psi_2, \Phi_1>-\frac 12 q<\Psi_1, \Phi_2>$$
$$+p<\Psi_1, \Phi_1>
-p<\Psi_2, \Phi_2>+\frac 12<\Psi_2, \Phi_1>^2
-\frac 14q^2<\Psi_2, \Phi_1>.$$\par
The Lax representations for the $x$-FDHS (2.24) and the
$t_1$-FDHS (2.26), which can can be deduced from the adjoint
representation
of (2.1) and (2.2), are given by (2.13) with  $\widetilde V^{(1)}$
defined
by
(2.25) and $\widetilde U$ obtained from $U$ by using $q$ instead of $u$
as
well as $M$ given by
$$A(\la)=p+\frac{1}{4}\sum_{j=1}^{N}\frac{\psi_{1j}\phi_{1j}-\psi_{2j}
\phi_{2j}}
{\la-\la_{j}},\qquad
B(\la)=\la-\frac 12q+\frac 12\sum_{j=1}^{N}\frac{\psi_{2j}\phi_{1j}}
{\la-\la_{j}},$$
$$C(\la)=-\la^2-\frac 12q\la+\frac 12<\Psi_2, \Phi_1>-\frac 14 q^2
+\frac 12\sum_{j=1}^{N}\frac{\psi_{1j}\phi_{2j}}{\la-\la_{j}}. \tag
2.27$$
\par
The  generating function of  integrals of motion for (2.24) and (2.26)
yields
$$A^2(\la)+B(\la)C(\la)\equiv P(\la)=-\la^3+P_0+
\sum_{j=1}^{N}[\frac{P_{j}}{\la-\la_{j}}
+\frac{P^2_{N+j}}{(\la-\la_{j})^2}], \tag 2.28$$
where $P_0,...,P_{2N}$ are independent integrals of
motion for the FDHSs
(2.24)
and (2.26) and $P_0=\frac 12F_1$,
$$P_j=-\frac 12\la_j^2\psi_{2j}\phi_{1j}+\frac 12\la_j
\psi_{1j}\phi_{2j}
-\frac 14\la_jq\psi_{2j}\phi_{1j}-\frac 14q\psi_{1j}\phi_{2j}$$
$$+\frac 12p(\psi_{1j}\phi_{1j}-\psi_{2j}\phi_{2j})
+\frac 14(<\Psi_2, \Phi_1>-\frac 12 q^2)\psi_{2j}\phi_{1j}$$
$$+\frac 18\sum_{k\neq j}\frac{1}{\la_j-\la_{k}}[(\psi_{1j}\phi_{1j}
-\psi_{2j}\phi_{2j})(\psi_{1k}\phi_{1k}
-\psi_{2k}\phi_{2k})+4\psi_{1j}\phi_{2j}\psi_{2k}
\phi_{1k}],\tag 2.29a$$
$$P_{N+j}=\frac 14(\psi_{1j}\phi_{1j}+\psi_{2j}\phi_{2j}),\qquad
j=1,...,N.\tag 2.29b$$
We have
$$F_1=2P_{0}, \qquad
F_2=2\sum_{j=1}^{N}P_{j}. \tag 2.30$$\par
Similarly, it can be shown that (2.24) and (2.26) are
FDIHSs and commute with each other. The KdV
equation (2.6) is factorized
by $x$-FDIHS (2.24) and $t_1$-FDIHS (2.26).
If $(\Psi_1, \Psi_2, p, \Phi_1, \Phi_2, q)$ satisfies the FDIHSs (2.24)
and (2.26)
simultaneously, then
$u=q$ solves the KdV equation (2.6).\par

\subhead {2.2 The separation of variables for the KdV equations}
\endsubhead\par
(1) For the case $k_0=0$, we first consider the separation of variables
for
FDIHSs (2.11) and (2.12).
With respect to the standard Poisson bracket (2.18), it is found that
for the $A(\la)$ and $B(\la)$ given by (2.14) we have
$$\{A(\la), A(\mu)\}=\{B(\la), B(\mu)\}=0,\quad
\{A(\la), B(\mu)\}=\frac 1{2(\la-\mu)}[B(\mu)-B(\la)].\tag 2.31$$
An effective way to introduce the separated variables $v_k, u_k$ and
to obtain the separated equations  is to use the Lax matrix $M$ and
the generating function of integrals of motion.
 The commutator relations
 (2.31) and the equation (2.15) enable
  us to
define the first $N$ pairs of the canonical variables $u_1,...,u_N$ by
the set of zeros of $B(\la)$ [1-3]
$$B(\la)=1+\frac 12\sum_{j=1}^{N}\frac{\psi_{2j}\phi_{1j}}
{\la-\la_{j}}=\frac {R(\la)}{K(\la)},\tag 2.32a$$
where
$$R(\la)=\prod_{k=1}^{N}(\la-u_{k}), \qquad
K(\la)=\prod_{k=1}^{N}(\la-\la_{k}), $$
and $v_1,...,v_N$ by
$$v_k=2A(u_k), \qquad k=1,...,N.\tag 2.32b$$
The commutator relations (2.31) guarantee that $u_1,...,u_N$ and
$v_1,...,v_N$ satisfy the canonical conditions (1.1) [1-3]. Then
substituting $u_k$ into (2.15) gives rise to the first $N$ separated
equations
$$v_k=2A(u_k)=2\sqrt{P(u_k)}
=2\sqrt{-u_k+\sum_{j=1}^{N}
[\frac{P_{j}}{u_k-\la_{j}}+\frac{P^2_{N+j}}{(u_k-\la_{j})^2}]}
,\quad k=1,...,N.\tag 2.33$$\par
The FDIHSs (2.11) and (2.12) have $2N$ degrees of freedom, we need
to introduce the other $N$ pairs of canonical variables
$v_k, u_k, k=N+1,...,2N$.
Notice that $P_{N+j}$ given by (2.16b) are integrals of motion for the
FDIHSs
 (2.11) and (2.12), and satisfy
$$\{B(\la), P_{N+j}\}=\{A(\la), P_{N+j}\}=0.\tag 2.34$$
Thus we may define
$$v_{N+j}=2P_{N+j}=\frac 12(\psi_{1j}\phi_{1j}+
\psi_{2j}\phi_{2j}),\quad
j=1,...,N,\tag 2.35a$$
which also give rise to the separated equations. It is easy to see that
if we take
$$u_{N+j}=\text {ln}\frac {\phi_{1j}}{\psi_{2j}}, \quad j=1,...,N,
\tag 2.35b$$
then
$$\{v_{N+j}, u_{N+k}\}=\delta_{jk},\quad\{v_{N+j}, v_{N+k}\}=
\{u_{N+j}, u_{N+k}\}=0,\quad j,k=1,...,N,\tag 2.36$$
$$\{B(\la), u_{N+j}\}=\{A(\la), u_{N+j}\}=\{B(\la), v_{N+j}\}=
\{A(\la), v_{N+j}\}=0.\tag 2.37$$

We have the following proposition.
\proclaim {Proposition 1}  Assume that $\la_j, \phi_{ij}, \psi_{ij}
\in\text {\bf R}, i=1,2, j=1,...,N$.
Introduce the separated variables
$u_{1},...,u_{2N}$
and  $v_{1},...,v_{2N}$ by (2.32) and (2.35).
If $u_1,...,u_N,$ are single zeros of $B(\la)$, then
$v_1,...,v_{2N}$ and $u_1,...,u_{2N}$ are canonically conjugated, i.e.,
they satisfy (1.1).\par
\endproclaim
{\it{Proof.}} By following the similar
method in [1-6,23,24], it is easy to show that
$v_1,...,v_{N}$ and $u_1,...,u_{N}$ satisfy (1.1).
Notice $B'(u_k)\neq 0$. Hereafter the prime denotes the differentiation
with respect to $\la$.
It follows from (2.36) and (2.37) that
$$0=\{u_{N+k},B(u_{l})\}=
B'(u_{l})\{u_{N+k}, u_{l}\}+
\{u_{N+k}, B(\mu)\}|_{\mu=u_{l}}=B'(u_{l})\{u_{N+k}, u_{l}\},$$
$$\{v_{k}, u_{N+l}\}
=2\{A(u_{k}), u_{N+l}\}$$
$$=2A'(u_k)\{u_k, u_{N+l}\}+\{A(\la), u_{N+l}\}|_{\la=u_{k}}=
2A'(u_k)\{u_k, u_{N+l}\}, \tag 2.38$$
which leads to $\{u_{N+k}, u_{l}\}=\{u_{N+k}, v_{l}\}=0$.
Similarly we can show that
$\{v_{N+k}, u_{l}\}=\{v_{N+k}, v_{l}\}=0.$
These together with (2.36) complete the proof. \par
It follows from (2.32a) and (2.35b) that
$$u=-<\Psi_{2}, \Phi_{1}>=2\sum_{j=1}^{N}(u_j-\la_j),\tag 2.39$$
$$ \psi_{2j}\phi_{1j}=2\frac{R(\la_j)}{K'(\la_{j})},\qquad
\frac {\phi_{1j}}{\psi_{2j}}=e^{u_{N+j}},\qquad j=1,...,N,$$
or
$$ \phi_{1j}=\sqrt{\frac{2R(\la_j)e^{u_{N+j}}}{K'(\la_{j})}},\qquad
\psi_{2j}=\sqrt{\frac{2R(\la_j)e^{-u_{N+j}}}{K'(\la_{j})}}
,\qquad j=1,...,N.\tag 2.40$$\par
The separated equations are given by (2.33) and (2.35a).
 Replacing $v_k$ by the partial derivative $\frac {\p S}{\p u_k}$ of
 the generating function $S$ of the canonical transformation and
 interpreting the $P_i$ as integration constants,
the equations (2.33) and (2.35a) give rise to the Hamilton-Jacobi
equations
which
are completely separable and may be integrated to give the completely
separated solution
$$S(u_1,...,u_{2N})=\sum_{k=1}^{N}[\int^{u_k}2\sqrt{P(\la)}d\la
+2P_{N+k}u_{N+k}]. \tag 2.41$$\par
The linearizing coordinates are then
$$Q_i=\frac {\p S}{\p P_i}=\sum_{k=1}^{N}\int^{u_k}
\frac {1}{(\la-\la_i)
\sqrt{ P(\la)}}d\la,
\quad i=1,...,N, \tag 2.42a$$
$$Q_{N+i}=\frac {\p S}{\p P_{N+i}}
=2\sum_{k=1}^{N}\int^{u_k}\frac {P_{N+i}}{(\la-\la_i)^2\sqrt {
P(\la)}}d\la
+2u_{N+i}, \quad i=1,...,N. \tag 2.42b$$\par
By using (2.17), the linear flow induced by (2.11) is then given by
$$Q_i=\gamma_i+x\frac {\p F_1}{\p P_i}=\gamma_i+2x,\quad
Q_{N+i}=2\gamma_{N+i}+x\frac {\p F_1}{\p P_{N+i}}=2\gamma_{N+i},
\quad i=1,...,N. \tag 2.43$$
Hereafter $\gamma_i, i=1,...,2N,$ are arbitrary constants. Combining
the equation (2.42) with the equation (2.43) leads to the Jacobi
inversion problem for the FDIHS (2.11)
$$\sum_{k=1}^{N}\int^{u_k}\frac {1}{(\la-\la_i)\sqrt {P(\la)}}d\la
=\gamma_{i}+2x,
\quad i=1,...,N, \tag 2.44a$$
$$\sum_{k=1}^{N}[\int^{u_k}\frac {P_{N+i}}{(\la-\la_i)^2\sqrt {
P(\la)}}d\la
+u_{N+i}=\gamma_{N+i}, \quad i=1,...,N. \tag 2.44b$$
If, by using the Jacobi inversion technique [7],
$\phi_{1j}, \psi_{2j},
<\Psi_{2}, \Phi_{1}>$ given by (2.39) and (2.40) can be obtained from
(2.44),
then $\phi_{2j}, \psi_{1j}$ can be found from the first and the last
equation in (2.11) by an algebraic calculation, respectively.
The $(\phi_{1j}, \phi_{2j}, \psi_{1j}, \psi_{2j})$ provides
the solution
to the FDIHS (2.11). \par
By using (2.17), the linear flow induced by (2.12) is then given by
$$Q_i=\bar\gamma_i+\frac {\p F_2}{\p
P_i}t_1=\bar\gamma_i+2\la_it_1,\quad $$
$$Q_{N+i}=2\bar\gamma_{N+i}+\frac {\p F_2}{\p
P_{N+i}}t_1=2\bar\gamma_{N+i}
+4P_{N+i}t_1,
\quad i=1,...,N, \tag 2.45$$
where $\bar\gamma_i$ are arbitrary constants.
Combining the equation (2.42) with the equation (2.45) leads to the
Jacobi
inversion problem for the FDIHS (2.12)
$$\sum_{k=1}^{N}\int^{u_k}\frac {1}{(\la-\la_i)\sqrt { P(\la)}}d\la
=\bar\gamma_i+2\la_it_1,\quad i=1,...,N, \tag 2.46a$$
$$\sum_{k=1}^{N}\int^{u_k}\frac {P_{N+i}}{(\la-\la_i)^2\sqrt {
P(\la)}}d\la
+u_{N+i}=\bar\gamma_{N+i}+2P_{N+i}t_1,
\quad i=1,...,N. \tag 2.46b$$\par
Finally, since the KdV equation (2.6) is factorized by the FDIHSs (2.11)

and (2.12), combining the equation (2.44) with the equation
 (2.46) give rise to the
 Jacobi inversion problem for the KdV equation (2.6)
$$\sum_{k=1}^{N}\int^{u_k}\frac {1}{(\la-\la_i)\sqrt {P(\la)}}d\la
=\gamma_{i}+2x
+2\la_it_1,\quad i=1,...,N, \tag 2.47a$$
$$\sum_{k=1}^{N}\int^{u_k}\frac {P_{N+i}}{(\la-\la_i)^2\sqrt {
P(\la)}}d\la
+u_{N+i}=\gamma_{N+i}+2P_{N+i}t_1,\quad i=1,...,N. \tag 2.47b$$
If, by using the  Jacobi inversion technique [7], $u$ given by (2.39)
can
be found in terms of Riemann theta functions by solving  (2.47),
then $u$ provides the solution of the KdV equation (2.6).\par
In general, since the n-th KdV equation (2.4) is factorized by the
$x$-FDIHS (2.11) and the $t_n$-FDIHS (2.22), the above procedure can
be applied to find the Jacobi inversion problem for the n-th KdV
equation (2.4). We have the following proposition.
\proclaim {Proposition 2} The Jacobi inversion problem for the n-th KdV
equation (2.4) is given by
$$\sum_{k=1}^{N}\int^{u_k}\frac {1}{(\la-\la_i)\sqrt {P(\la)}}d\la
=\gamma_{i}+2x$$
$$+t_n\sum_{m=0}^{n}(\frac 12)^{m-1}\alpha_m\sum_{l_1+...+l_{m+1}=n+2}
\la_i^{l_{m+1}-2}\widetilde F_{l_1}...\widetilde F_{l_{m}},
\quad i=1,...,N, $$
$$\sum_{k=1}^{N}\int^{u_k}\frac {P_{N+i}}{(\la-\la_i)^2
\sqrt {P(\la)}}d\la
+u_{N+i}=\gamma_{N+i}$$
$$+t_n\sum_{m=0}^{n}(\frac 12)^{m-2}\alpha_m\sum_{l_1+...+l_{m+1}=n+2}
(l_{m+1}
-2)\la_i^{l_{m+1}-3}P_{N+i}\widetilde F_{l_1}...\widetilde F_{l_{m}},
 \quad i=1,...,N, $$
where $l_1\geq 1,...,l_{m+1}\geq 1$ and $\widetilde F_{l_1},...
\widetilde F_{l_{m}},$ are given by (2.21).
\endproclaim\par
(2) For the case $k_0=1$, we now consider the separation of variables
for
FDIHSs (2.24) and (2.26).
With respect to the standard Poisson bracket, it is found that
 the $A(\la)$ and $B(\la)$ given by (2.27) also satisfy commutator
relation
 (2.31).
In the same way,
the first $N+1$ pairs of the canonical variables $u_1,...,u_{N+1}$ can
be introduced by
the set of zeros of $B(\la)$
$$B(\la)=\la-\frac 12q+\frac 12\sum_{j=1}^{N}\frac{\psi_{2j}\phi_{1j}}
{\la-\la_{j}}=\frac {R(\la)}{K(\la)},\tag 2.48a$$
where
$$R(\la)=\prod_{k=1}^{N+1}(\la-u_{k}), \qquad
K(\la)=\prod_{k=1}^{N}(\la-\la_{k}), $$
and $v_1,...,v_{N+1}$ by
$$v_k=2A(u_k), \qquad k=1,...,N+1.\tag 2.48b$$
Then substituting $u_k$ into (2.28) gives rise to the first $N+1$
separated
equations
$$v_k=2A(u_k)=2\sqrt {P(u_k)}
=2\sqrt {-u_k^3+P_0+\sum_{j=1}^{N}[\frac{P_{j}}{u_k-\la_{j}}
+\frac{P^2_{N+j}}{(u_k-\la_{j})^2}]},$$
$$\qquad\quad k=1,...,N+1.\tag 2.49$$\par
The additional $N$ pairs of canonical variables can also be defined by
the
same way
$$v_{N+1+j}=2P_{N+j}=\frac
12(\psi_{1j}\phi_{1j}+\psi_{2j}\phi_{2j}),\quad
j=1,...,N,\tag 2.50a$$
$$u_{N+1+j}=\text {ln}\frac {\phi_{1j}}{\psi_{2j}}, \quad
 j=1,...,N.\tag 2.50b$$\par
In the same way we can show the following proposition.
\proclaim {Proposition 3}  Assume that $\la_j, \phi_{ij}, \psi_{ij}
\in\text {\bf R}, i=1,2, j=1,...,N$.
Introduce the separated variables
$u_{1},...,u_{2N+1}$
and  $v_{1},...,v_{2N+1}$ by (2.48) and (2.50).
If $u_1,...,u_{N+1},$ are single zeros of $B(\la)$, then
$v_1,...,v_{2N+1}$ and $u_1,...,u_{2N+1}$ are canonically conjugated,
i.e.,
they satisfy (1.1).\par
\endproclaim\par
It follows from (2.48) and (2.50) that
$$u=q=2\sum_{j=1}^{N+1}u_j-2\sum_{j=1}^{N}\la_j,\tag 2.51a$$
$$ \phi_{1j}=\sqrt{\frac{2R(\la_j)e^{u_{N+1+j}}}{K'(\la_{j})}},\qquad
\psi_{2j}=\sqrt{\frac{2R(\la_j)e^{-u_{N+1+j}}}{K'(\la_{j})}}
,\qquad j=1,...,N.\tag 2.51b$$\par
The separated equations  (2.49) and (2.50a)
 may be integrated to give the completely separated solution for the
 generating function $S$ of the canonical transformation
$$S(u_1,...,u_{2N+1})=\sum_{k=1}^{N+1}\int^{u_k}2\sqrt {P(\la)}d\la
+2\sum_{k=1}^{N}P_{N+k}u_{N+1+k}, \tag 2.52$$
where $P(\la)$ is given by (2.28).\par
In the exactly same way, one gets the
Jacobi inversion problem for the FDIHS (2.24)
$$\sum_{k=1}^{N+1}\int^{u_k}\frac {1}{\sqrt {P(\la)}}d\la
=\gamma_{0}+2x, \tag 2.53a$$
$$\sum_{k=1}^{N+1}\int^{u_k}\frac {1}{(\la-\la_i)\sqrt {P(\la)}}d\la
=\gamma_{i},\quad i=1,...,N, \tag 2.53b$$
$$\sum_{k=1}^{N+1}[\int^{u_k}\frac {P_{N+i}}{(\la-\la_i)^2
\sqrt { P(\la)}}d\la
+u_{N+1+i}=\gamma_{N+i}, \quad i=1,...,N, \tag 2.53c$$
the Jacobi inversion problem for the FDIHS (2.26)
$$\sum_{k=1}^{N+1}\int^{u_k}\frac {1}{\sqrt {P(\la)}}d\la
=\gamma_{0}, \tag 2.54a$$
$$\sum_{k=1}^{N+1}\int^{u_k}\frac {1}{(\la-\la_i)\sqrt { P(\la)}}d\la
=\gamma_i+2t_1,\quad i=1,...,N, \tag 2.54b$$
$$\sum_{k=1}^{N+1}\int^{u_k}\frac {P_{N+i}}{(\la-\la_i)^2\sqrt {
P(\la)}}d
\la
+u_{N+1+i}=\gamma_{N+i},\quad i=1,...,N. \tag 2.54c$$\par
Finally we have the following proposition.
\proclaim {Proposition 4}
 The Jacobi inversion problem for the KdV equation (2.6)
$$\sum_{k=1}^{N+1}\int^{u_k}\frac {1}{\sqrt {P(\la)}}d\la
=\gamma_{0}+2x, \tag 2.55a$$
$$\sum_{k=1}^{N+1}\int^{u_k}\frac {1}{(\la-\la_i)\sqrt {P(\la)}}d\la
=\gamma_{i}
+2t_1,\quad i=1,...,N, \tag 2.55b$$
$$\sum_{k=1}^{N+1}\int^{u_k}\frac {P_{N+i}}{(\la-\la_i)^2\sqrt {
P(\la)}}d
\la
+u_{N+1+i}=\gamma_{N+i},\quad i=1,...,N. \tag 2.55c$$\endproclaim\par
If, by using the  Jacobi inversion technique [7], $u$ given by (2.51a)
can
be found in terms of Riemann theta functions by solving  (2.55),
then $u$ provides the solution of the KdV equation (2.6).\par
In general, since the n-th KdV equation (2.4) is factorized by the
$x$-FDIHS (2.24) and the $t_n$-FDIHS obtained from (2.9) under (2.24),
the above procedure can
be applied to find the Jacobi inversion problem for the n-th KdV
equation (2.4).\par
(3) The method can be applied to all high-order binary constrained flows

(2.8) and (2.9) as well as the whole KdV hierarchy. For any fixed $k_0$,

by
introducing the so-called Jacobi-Ostrogradsky coordinates, the
high-order
binary
$x$-constrained flow (2.8) can be transformed into a $x$-FDIHS with
degree
of freedom
$2N+k_0$. Under the $x$-FDIHS, the  binary $t_n$-constrained flow (2.9)
can
also be transformed into a $t_n$-FDIHS. The Lax representation for the
$x$-
and $t_n$-FDIHS can be deduced from the adjoint representation of (2.1)
and (2.2) by using the method in [27,28]. By means of the Lax matrix we
can
introduce the first $N+k_0$ canonical variables $u_1,...,u_{N+k_0}$
by the set of zeros of $B(\la)$ and $v_k=2A(u_k), k=1,...,N+k_0$. Then
 the additional $N$ canonical separated variables can be defined by
$$v_{N+k_0+j}=2P_{N+j}=\frac
12(\psi_{1j}\phi_{1j}+\psi_{2j}\phi_{2j}),\quad
u_{N+k_0+j}=\text {ln}\frac {\phi_{1j}}{\psi_{2j}}, \quad
 j=1,...,N.$$
Finally, since the n-th KdV equation (2.4) is factorized by the
$x$-FDIHS and the $t_n$-FDIHS,
in the exactly same way we can obtain the Jacobi inversion problem for
(2.4). The above scheme can be applied to all soliton
equations admitting $2\times 2$ Lax pairs.\par

\subhead {3. The separation of variables for the AKNS equations}
\endsubhead\par

\subhead {3.1 Binary constrained flows of the AKNS
hierarchy}\endsubhead\par

For the AKNS spectral problem [30]
$$\phi_x=U(u,\la)\phi,\quad
 U(u, \lambda)
=\left( \matrix -\la&q\\r&\la\endmatrix\right),\quad
\phi=\binom {\phi_{1}}{\phi_{2}},\quad u=\binom {q}{r},\tag 3.1$$
Take
$$\phi_{t_n}=V^{(n)}(u,\la)\phi,\qquad V^{(n)}(u, \la)=\sum_{i=0}^{n}
\left( \matrix a_i&b_i\\c_i&-a_i\endmatrix\right)\la^{n-i},\tag 3.2$$
where
$$a_{0}=-1,\quad b_{0}=c_{0}=0,\quad a_1=0,\quad b_{1}=q,\quad
c_{1}=r,\quad a_2=\frac{1}{2}qr,...,$$
$$\binom{c_{k+1}}{ b_{k+1
}}=L \binom{c_{k}}{ b_{k}}, \qquad
a_k=\p^{-1}(qc_k-rb_k),\qquad
 k=1,2,\cdots,\tag 3.3$$
$$L=\frac 12\left( \matrix \p-2r\p^{-1}q&2r\p^{-1}r\\-2q\p^{-1}q&-\p
+2q\p^{-1}r\endmatrix\right).$$\par
The AKNS hierarchy associated with (3.1) and (3.2) reads [30]
$$u_{t_n}={\binom {q}{r}}_{t_n}
=JL^n\binom {r}{q}
=J\frac {\delta H_{n+1}}{\delta u},\qquad n=1,2,\hdots, \tag 3.4$$
$$J=\left( \matrix 0&-2\\2&0\endmatrix\right),\qquad
H_{n}=\frac{2a_{n+1}}{n+1},\qquad
\binom {c_n}{b_n}
=\frac {\delta H_{n}}{\delta u},\quad n=1,2,\hdots.$$\par
For $n=2$ we have
$$\phi_{t_2}=V^{(2)}(u,\la)\phi,\qquad
V^{(2)}=\left( \matrix -\la^2+\frac 12qr&q\la-\frac 12q_x\\r\la+
\frac 12r_x&
\la^2-\frac 12qr\endmatrix\right),\tag 3.5$$
and the AKNS equation (3.4) for $n=2$ reads
$$q_{t_2}=-\frac 12q_{xx}+q^2r, \qquad r_{t_2}=\frac 12r_{xx}-r^2q.
\tag 3.6$$\par
The adjoint AKNS spectral problem is of the same form as equation (2.7).

We have
$$\frac {\delta\la}{\delta u}=\binom{\frac {\delta\la}{\delta q}}
{\frac {\delta\la}{\delta r}}=Tr[\left(\matrix \phi_1\psi_1
& \phi_1\psi_2\\ \phi_2\psi_1&\phi_2\psi_2
\endmatrix\right)\frac {\p U(u,\la)}{\p u}]
=\binom {\psi_1\phi_2}{\psi_2\phi_1},\tag 3.7$$
which should be read componentwise [26].
\par
The binary $x$-constrained flows of the AKNS
hierarchy (3.4) are defined
by [15,17,21]
$$
\Phi_{1,x}=-\La\Phi_{1}+q\Phi_{2},\qquad\Phi_{2,x}
=r\Phi_{1}+\La\Phi_{2}
,\tag 3.8a$$
$$
\Psi_{1,x}=\La\Psi_{1}-r\Psi_{2},\qquad\Psi_{2,x}
=-q\Psi_{1}-\La\Psi_{2},
\tag 3.8b$$
$$\frac {\delta H_{k_0+1}}{\delta u}-
\sum_{j=1}^{N}\frac {\delta \lambda_{j}}{\delta u}
=\binom {c_{k_0+1}}{ b_{k_0+1}}-\beta\binom {
<\Psi_1,\Phi_2>}{ <\Psi_2,\Phi_1>}=0.\tag 3.8c$$\par
(1) For $k_0=0, \beta=1$, we have
$$\binom {c_{1}}{ b_{1}}=\binom {r}{q}=\binom {
<\Psi_1,\Phi_2>}{ <\Psi_2,\Phi_1>}.\tag 3.9$$
By substituting (3.9) into (3.8a) and (3.8b), one gets a $x$-FDHS
[15,17]
$$ \Phi_{1x}=\frac {\p F_1}{\p \Psi_1},\quad \Phi_{2x}=\frac {\p F_1}
{\p \Psi_2},\quad
 \Psi_{1x}=-\frac {\p F_1}{\p \Phi_1},\quad
 \Psi_{2x}=-\frac {\p F_1}{\p \Phi_2},\tag 3.10$$
$$F_1=<\La\Psi_2, \Phi_2>-<\La\Psi_1, \Phi_1>
+<\Psi_2, \Phi_1><\Psi_1, \Phi_2>.$$\par
Under the constraint (3.9) and the FDHS (3.10), the binary
$t_2$-constrained
flow obtained from  (3.2) with $V^{(2)}$ given by (3.5) and its adjoint
equation for $N$ distinct real number $\la_j$ can also be written
as a $t_2$-FDHS
$$ \Phi_{1,t_2}=\frac {\p F_2}{\p \Psi_1},\quad \Phi_{2,t_2}=
\frac {\p F_2}{\p \Psi_2},\quad
 \Psi_{1,t_2}=-\frac {\p F_2}{\p \Phi_1},\quad
 \Psi_{2,t_2}=-\frac {\p F_2}{\p \Phi_2},\tag 3.11$$
$$F_2=<\La^2\Psi_2, \Phi_2>-<\La^2\Psi_1, \Phi_1>
+<\Psi_2, \Phi_1><\La \Psi_1, \Phi_2>$$
$$+<\La \Psi_2, \Phi_1><\Psi_1, \Phi_2>
-\frac 12(<\Psi_2, \Phi_2>-<\Psi_1, \Phi_1>)<\Psi_2, \Phi_1>
<\Psi_1, \Phi_2>.$$\par
The Lax representation for the FDHSs (3.10) and (3.11) which can also
be deduced from the adjoint representation of (3.1) and (3.2) are
presented
by (2.13) with the entries of the
 Lax matrix $M$ given by [21]
$$A(\la)=-1+\frac{1}{2}\sum_{j=1}^{N}\frac{\psi_{1j}\phi_{1j}-\psi_{2j}
\phi_{2j}}
{\la-\la_{j}},\tag 3.12a$$
$$B(\la)=\sum_{j=1}^{N}\frac{\psi_{2j}\phi_{1j}}{\la-\la_{j}},\qquad
C(\la)=\sum_{j=1}^{N}\frac{\psi_{1j}\phi_{2j}}{\la-\la_{j}}. \tag
3.12b$$
A straightforward calculation yields
$$A^2(\la)+B(\la)C(\la)\equiv P(\la)=1+
\sum_{j=1}^{N}[\frac{P_{j}}{\la-\la_{j}}+\frac{P^2_{N+j}}
{(\la-\la_{j})^2}],
\tag 3.13$$
where $P_1,...,P_{2N}$ are independent integrals of motion for the FDHSs

(3.10) and (3.11)
$$P_j=\psi_{2j}\phi_{2j}-\psi_{1j}\phi_{1j}$$
$$+\frac 12\sum_{k\neq j}\frac{1}{\la_j-\la_{k}}[(\psi_{1j}\phi_{1j}-
\psi_{2j}\phi_{2j})(\psi_{1k}\phi_{1k}
-\psi_{2k}\phi_{2k})
+4\psi_{1j}\phi_{2j}\psi_{2k}\phi_{1k}],\tag 3.14a$$
$$P_{N+j}=\frac 12(\psi_{1j}\phi_{1j}+\psi_{2j}\phi_{2j}),\qquad
j=1,...,N.\tag 3.14b$$
It is easy to verify that
$$F_1=\sum_{j=1}^{N}(\la_jP_{j}+P^2_{N+j})-\frac
14(\sum_{j=1}^{N}P_{j})^2,
\tag 3.15a$$
$$F_2=\sum_{j=1}^{N}(\la_j^2P_{j}+2\la_jP^2_{N+j})-\frac 12(
\sum_{j=1}^{N}P_{j})
\sum_{j=1}^{N}(\la_jP_{j}+P^2_{N+j})+\frac 18(\sum_{j=1}^{N}P_{j})^3.
\tag 3.15b$$\par
Similarly, it is easy to show that $P_1,...,P_{2N}$ are in involution,
(3.10) and (3.11)
 are FDIHSs. The AKNS equation (3.6)
is factorized by the $x$-FDIHS (3.10) and the $t_2$-FDIHS (3.11),
namely,
if $(\Psi_1, \Psi_2, \Phi_1,
\Phi_2)$ satisfies the FDIHSs (3.10) and (3.11) simultaneously, then
$(q, r)$ given by (3.9) solves the AKNS equation (3.6). In general,
the factorization of the n-th AKNS equations (3.4) will be presented
in the end of section 3.2.\par

(2) For $k_0=1, \beta=\frac 12$, (3.8c) yields
$$r_x=<\Psi_1,\Phi_2>,\quad q_x=-<\Psi_2,\Phi_1>.\tag 3.16$$
The equations (3.8a), (3.8b) and (3.16) can be written as a $x$-FDIHS
$$ \Phi_{ix}=\frac {\p F_1}{\p \Psi_1},\quad r_{x}
=\frac {\p F_1}{\p q},
\quad
 \Psi_{ix}=-\frac {\p F_1}{\p \Phi_i},\quad
q_{x}=-\frac {\p F_1}{\p r},\quad i=1,2,\tag 3.17$$
$$F_1=<\La\Psi_2, \Phi_2>-<\La\Psi_1, \Phi_1>
+r<\Psi_2, \Phi_1>+q<\Psi_1, \Phi_2>.$$\par
Under the constraint (3.16) and the FDIHS (3.17),
$V^{(2)}$ becomes
$$\widetilde V^{(2)}=\left( \matrix -\la^2+\frac 12qr&q\la
+\frac 12<\Psi_2, \Phi_1>\\r\la+\frac 12<\Psi_1, \Phi_2>&
\la^2-\frac 12qr\endmatrix\right).\tag 3.18$$
Then under the constraint (3.16) and the FDIHS (3.17),
the binary $t_2$-constrained consisting of replicas (3.5) and its
adjoint
system
for $N$ distinct real number $\la_j$ as well as (3.6) can also be
written
as a $t_2$-FDIHS
$$ \Phi_{i,t_2}=\frac {\p F_2}{\p \Psi_i},\quad r_{t_2}=\frac {\p
F_2}{\p q},
\quad
 \Psi_{i,t_2}=-\frac {\p F_2}{\p \Phi_i},\quad
 q_{t_2}=-\frac {\p F_2}{\p r} \quad i=1,2,\tag 3.19$$
$$F_2=<\La^2\Psi_2, \Phi_2>-<\La^2\Psi_1, \Phi_1>
+q<\La \Psi_1, \Phi_2>+r<\La \Psi_2, \Phi_1>$$
$$-\frac 12qr(<\Psi_2, \Phi_2>-<\Psi_1, \Phi_1>)
+\frac 12<\Psi_2, \Phi_1><\Psi_1, \Phi_2>-\frac 12q^2r^2.$$\par
The Lax matrix $M$ for FDIHS (3.17) and (3.19) is given by
$$A(\la)=-\la+\frac{1}{4}\sum_{j=1}^{N}\frac{\psi_{1j}
\phi_{1j}-\psi_{2j}\phi_{2j}}
{\la-\la_{j}},\tag 3.20a$$
$$B(\la)=q+\frac 12\sum_{j=1}^{N}
\frac{\psi_{2j}\phi_{1j}}{\la-\la_{j}},
\qquad
C(\la)=r+\frac 12\sum_{j=1}^{N}\frac{\psi_{1j}\phi_{2j}}{\la-\la_{j}}.
\tag 3.20b$$
A straightforward calculation yields
$$A^2(\la)+B(\la)C(\la)\equiv P(\la)=\la^2+P_0
+\sum_{j=1}^{N}[\frac{P_{j}}{\la-\la_{j}}
+\frac{P^2_{N+j}}{(\la-\la_{j})^2}], \tag 3.21$$
where $P_0,...,P_{2N}$ are independent integrals of motion
in involution
for the FDIHSs (3.17) and (3.19)
$$P_0=\frac 12(<\Psi_2, \Phi_2>-<\Psi_1, \Phi_1>)+qr,$$
$$P_j=\frac 12[\la_j\psi_{2j}\phi_{2j}-\la_j\psi_{1j}\phi_{1j}
+q\psi_{1j}\phi_{2j}+r\psi_{2j}\phi_{1j}]$$
$$+\frac 18\sum_{k\neq j}\frac{1}{\la_j-\la_{k}}[(\psi_{1j}\phi_{1j}-
\psi_{2j}\phi_{2j})(\psi_{1k}\phi_{1k}
-\psi_{2k}\phi_{2k})
+4\psi_{1j}\phi_{2j}\psi_{2k}\phi_{1k}],$$
$$P_{N+j}=\frac 14(\psi_{1j}\phi_{1j}+\psi_{2j}\phi_{2j}),
\qquad j=1,...,N.$$
It is easy to verify that
$$F_1=2\sum_{j=1}^{N}P_{j},\quad
F_2=2\sum_{j=1}^{N}(\la_jP_{j}+P^2_{N+j})-\frac 12P_{0}^2.\tag
3.22$$\par
It is easy to show that $P_1,...,P_{2N}$ are in involution, (3.17)
and (3.18)
 are FDIHSs and commute each other. The AKNS equation (3.6)
is factorized by the $x$-FDIHS (3.17) and the $t_2$-FDIHS (3.19),
namely,
if $(\Psi_1, \Psi_2, q, \Phi_1,\Phi_2, r)$ satisfies the FDIHSs (3.17)
and (3.19) simultaneously, then $(q, r)$ solves the AKNS equation (3.6).

\par

\subhead {3.2 The separation of variables for the AKNS equations }
\endsubhead\par
(1) For $k_0=0$ case, we present the Jacobi inversion problem
for (3.10)
and
(3.11) as well as for (3.6).
With respect to the standard Poisson bracket, $A(\la)$ and $B(\la)$
given
by (3.12) satisfy
$$\{A(\la), A(\mu)\}=\{B(\la), B(\mu)\}=0,\quad
\{A(\la), B(\mu)\}=\frac 1{\la-\mu}[B(\mu)-B(\la)].\tag 3.23$$\par
In contrast with the $B(\la)$  for the constrained
KdV flows, the $B(\la)$ given by (3.12b) has only $N-1$ zeros,
one has to define the canonical variables $u_k, v_k, k=1,...,N,$ in a
little different way:
$$B(\la)=\sum_{j=1}^{N}\frac{\psi_{2j}\phi_{1j}}
{\la-\la_{j}}=e^{u_N}\frac {R(\la)}{K(\la)},\quad
R(\la)=\prod_{k=1}^{N-1}(\la-u_{k}),\quad
K(\la)=\prod_{k=1}^{N}(\la-\la_{k}),\tag 3.24a$$
$$v_k= A(u_k), \quad k=1,...,N-1,\quad
v_N=\frac 12(<\Psi_{1}, \Phi_{1}>-<\Psi_{2}, \Phi_{2}>).\tag 3.24b$$
The equation (3.24a) yields
$$u_N=\text {ln}<\Psi_{2}, \Phi_{1}>.\tag 3.24c$$
Then it is easy to verify that
$$\{u_N, B(\mu)\}=\{v_N, A(\mu)\}=0,\qquad \{v_N, u_N\}=1,\tag 3.25a$$
$$\{u_N, A(\mu)\}=-\frac {B(\mu)}{<\Psi_{2}, \Phi_{1}>},\qquad
\{v_N, B(\mu)\}=B(\mu).\tag 3.25b$$
The commutator relations (3.23) and (3.25) guarantee that
$u_1,...,u_N,$
$ v_1,..., v_N$ satisfy the canonical conditions (1.1).
Similarly, we define
$$v_{N+j}=P_{N+j}, \quad u_{N+j}=\text {ln}
\frac {\phi_{1j}}{\psi_{2j}},
\quad j=1,...,N. \tag 3.26$$\par
In the same way we can show the following proposition.
\proclaim {Proposition 5}
Assume that  $\la_j, \phi_{ij}, \psi_{ij} \in\text {\bf R}, i=1,2,
j=1,...,N$.
Introduce the separated variables
$u_{1},...,u_{2N}$
and  $v_{1},...,v_{2N}$ by (3.24) and (3.26).
If $u_1,...,u_{N-1},$ are single zeros of $B(\la)$, then
$v_1,...,v_{2N}$ and $u_1,...,u_{2N}$ are canonically conjugated, i.e.,
they satisfy (1.1).\par
\endproclaim\par
It follows from (3.24) that
$$q=e^{u_{N}},\tag 3.27$$
$$ \psi_{2j}\phi_{1j}=e^{u_{N}}\frac{R(\la_j)}{K'(\la_{j})},\qquad
\frac {\phi_{1j}}{\psi_{2j}}=e^{u_{N+j}},\qquad j=1,...,N,$$
or
$$ \phi_{1j}=\sqrt{\frac{e^{u_{N}+u_{N+j}}
R(\la_j)}{K'(\la_{j})}},\qquad
\psi_{2j}=\sqrt{\frac{e^{u_{N}-u_{N+j}}
R(\la_j)}{K'(\la_{j})}},\qquad
 j=1,...,N.\tag 3.28$$
It is easy to see from (3.13) that
$$v_N=\frac 12(<\Psi_{1}, \Phi_{1}>-<\Psi_{2}, \Phi_{2}>)
=-\frac 12\sum_{i=1}^{N}P_{i}. \tag 3.29$$
 Then the separated equations obtained by substituting
$u_k$ into (3.13)
 and using (3.24) and the separated equations (3.26) and (3.29) may be
 integrated to give the generating function of the canonical
transformation
$$S(u_1,...,u_{2N})=\sum_{k=1}^{N-1}\int^{u_k}\sqrt {P(\la)}d\la
-\frac 12\sum_{i=1}^{N}P_iu_N+\sum_{i=1}^{N}P_{N+i} u_{N+i}.
 \tag 3.30$$\par
The linearizing coordinates are then
$$Q_i=\frac {\p S}{\p P_i}=\frac 12\sum_{k=1}^{N-1}\int^{u_k}
\frac{1}{(\la
-\la_i)\sqrt { P(\la)}}d\la-\frac 12u_{N}, \quad i=1,...,N, \tag 3.31a$$

$$Q_{N+i}=\frac {\p S}{\p P_{N+i}}
=\sum_{k=1}^{N-1}\int^{u_k}\frac {P_{N+i}}{(\la-\la_i)^2\sqrt {
P(\la)}}d\la
+u_{N+i},
 \quad i=1,...,N. \tag 3.31b$$\par
By using (3.15a), the linear flow induced by the FDIHS (3.10)
together with the equations (3.31) leads to the Jacobi inversion problem

for the FDIHS (3.10)
$$\sum_{k=1}^{N-1}\int^{u_k}\frac {1}{(\la-\la_i)\sqrt{P(\la)}}
d\la-u_{N}
=\gamma_{i}+(2\la_i-\sum_{k=1}^{N}P_k)x,
\quad i=1,...,N, \tag 3.32a$$
$$\sum_{k=1}^{N-1}\int^{u_k}\frac{P_{N+i}}{(\la-\la_i)^2
\sqrt{P(\la)}}d\la
+u_{N+i}
=\gamma_{N+i}+2P_{N+i}x,\quad i=1,...,N. \tag 3.32b$$\par
By using (3.15b), the linear flow induced by the FDIHS (3.11)
and the equations (3.31) result in the Jacobi inversion problem for
the FDIHS (3.11)
$$\sum_{k=1}^{N-1}\int^{u_k}\frac{1}{(\la-\la_i)\sqrt{ P(\la)}}
d\la-u_{N}$$
$$=\bar\gamma_i+[2\la_i^2
-\sum_{k=1}^{N}(\la_kP_k+\la_iP_k+P^2_{N+k})
+\frac 34(\sum_{k=1}^{N}P_k)^2]t_2,
\quad i=1,...,N, \tag 3.33a$$
$$\sum_{k=1}^{N-1}\int^{u_k}\frac{P_{N+i}}{(\la-\la_i)^2
\sqrt{P(\la)}}d\la+u_{N+i}
=\bar\gamma_{N+i}+P_{N+i}(4\la_i-\sum_{k=1}^{N}P_k)t_2,\quad i=1,...,N.
\tag 3.33b$$\par
Then, since the AKNS equations (3.6) are factorized by the FDIHSs (3.10)

and (3.11),  combining the equations (3.32) with the equations (3.33)
gives rise to the
 Jacobi inversion problem for the AKNS equations (3.6)
$$\sum_{k=1}^{N-1}\int^{u_k}\frac {1}{(\la-\la_i)\sqrt
{P(\la)}}d\la-u_{N}
$$
$$=\gamma_{i}+(2\la_i-\sum_{k=1}^{N}P_k)x
+[2\la_i^2
-\sum_{k=1}^{N}(\la_kP_k+\la_iP_k+P^2_{N+k})
+\frac 34(\sum_{k=1}^{N}P_k)^2]t_2,\tag 3.34a$$
$$\sum_{k=1}^{N}\int^{u_k}\frac {P_{N+i}}{(\la-\la_i)^2\sqrt {
P(\la)}}d\la
+u_{N+i}
=\gamma_{N+i}+2P_{N+i}x+P_{N+i}(4\la_i-\sum_{k=1}^{N}P_k)t_2,$$
$$\qquad\quad i=1,...,N. \tag 3.34b$$
If $\phi_{1j}, \psi_{2j}, q$ defined by (3.27) and (3.28) can be solved
from (3.34)
by using the Jacobi inversion technique, then $\phi_{2j}, \psi_{1j}$ can

be obtained from the first equation  and the last equation in (3.10) by
an
algebraic calculation, respectively. Finally $q$ and $r=<\Psi_1,\Phi_2>$

provides the solution to the AKNS equations (3.6).\par
Comparing (2.20) with (3.13), one gets
$$\widetilde F_0=1,\quad \widetilde F_k=\sum_{j=1}^{N}[\la_j^{k-1}P_j
+(k-1)\la_j^{k-2}P_{N+j}^2],\quad k=1,2,..., \tag 3.35$$
where $\widetilde F_k, k=1,2,...,$ are also integrals of motion
for both the FDIHS (3.10) and the $t_n$-binary constrained flow.
The n-th AKNS equations (3.4) are factorized by the $x$-FDIHS (3.10)
and the $t_n$-FDIHS with the Hamiltonian $F_n$ given by
$$F_n=2\sum_{m=0}^{n}(-\frac 12)^{m}\frac {\alpha_m}{m+1}
\sum_{l_1+...+l_{m+1}
=n+1}\widetilde F_{l_1}...\widetilde F_{l_{m+1}}, \tag 3.36$$
where $l_1\geq 1,...,l_{m+1}\geq 1, \alpha_m$ are given by (2.22c).
We have the proposition:
\proclaim {Proposition 6} The Jacobi inversion problem for the n-th AKNS

equations (3.4) is
$$\sum_{k=1}^{N-1}\int^{u_k}\frac {1}{(\la-\la_i)\sqrt
{P(\la)}}d\la-u_{N}
=\gamma_{i}+(2\la_i-\sum_{k=1}^{N}P_k)x$$
$$+2t_n\sum_{m=0}^{n}(-\frac 12)^m\alpha_m\sum_{l_1+...+l_{m+1}=n+1}
\la_i^{l_{m+1}-1}\widetilde F_{l_1}...\widetilde F_{l_{m}},
\quad i=1,...,N, $$
$$\sum_{k=1}^{N}[\int^{u_k}\frac {P_{N+i}}{(\la-\la_i)^2\sqrt {
P(\la)}}d\la
+u_{N+i}
=\gamma_{N+i}+2P_{N+i}x$$
$$+4t_n\sum_{m=0}^{n}(-\frac 12)^m\alpha_m\sum_{l_1+...+l_{m+1}=n+1}
(l_{m+1}-1)\la_i^{l_{m+1}-2}P_{N+i}\widetilde F_{l_1}...
\widetilde F_{l_{m}},
\quad i=1,...,N, $$
where $l_1\geq 1,...,l_{m+1}\geq 1,$ and $
\widetilde F_{l_1},...\widetilde F_{l_{m}},$ are given by (3.35).
\endproclaim\par
(2) For $k_0=1$ case,
with respect to the standard Poisson bracket, $A(\la)$ and $B(\la)$
given
by (3.20) also satisfy the commutator relation (2.31).
One  define the first $N+1$ pair of canonical variables $u_k, v_k,
k=1,...,N+1,$ in the following way:
$$B(\la)=q+\frac 12\sum_{j=1}^{N}\frac{\psi_{2j}\phi_{1j}}
{\la-\la_{j}}=e^{u_N+1}\frac {R(\la)}{K(\la)},\tag 3.37a$$
with
$$R(\la)=\prod_{k=1}^{N}(\la-u_{k}),\qquad
K(\la)=\prod_{k=1}^{N}(\la-\la_{k}),$$
and
$$v_k=2A(u_k), \quad k=1,...,N,\tag 3.37b$$
$$v_{N+1}=P_0=qr-\frac 12(<\Psi_{1}, \Phi_{1}>-<\Psi_{2}, \Phi_{2}>).
\tag 3.37c$$
The equation (3.24a) yields
$$u_{N+1}=\text {ln}q.\tag 3.37d$$
Then it is easy to verify that
$$\{u_{N+1}, B(\mu)\}=\{v_{N+1}, A(\mu)\}=0,\qquad \{v_{N+1},
u_{N+1}\}=1,$$
$$\{u_{N+1}, A(\mu)\}=0,\qquad
\{v_{N+1}, B(\mu)\}=B(\mu).\tag 3.38$$
Similarly, we define
$$v_{N+1+j}=2P_{N+j}, \qquad j=1,...,N,\tag 3.39a$$
$$ u_{N+1+j}=\text {ln}\frac {\phi_{1j}}{\psi_{2j}},
\quad j=1,...,N. \tag 3.39b$$  \par
In the same way we can show the following proposition.
\proclaim {Proposition 7}
Assume that  $\la_j, \phi_{ij}, \psi_{ij} \in\text {\bf R}, i=1,2,
j=1,...,N$.
Introduce the separated variables
$u_{1},...,u_{2N+1}$
and  $v_{1},...,v_{2N+1}$ by (3.37) and (3.39).
If $u_1,...,u_N,$ are single zeros of $B(\la)$, then
$v_1,...,v_{2N+1}$ and $u_1,...,u_{2N+1}$ are canonically conjugated,
i.e., they satisfy (1.1).\par
\endproclaim\par
It follows from (3.37) that
$$q=e^{u_{N+1}},\tag 3.40a$$
$$ \phi_{1j}=\sqrt{\frac{2e^{u_{N+1}+u_{N+1+j}}R(\la_j)}{K'(\la_{j})}},
\qquad
\psi_{2j}=\sqrt{\frac{2e^{u_{N+1}-u_{N+1+j}}
R(\la_j)}{K'(\la_{j})}},\qquad
 j=1,...,N.\tag 3.40b$$\par
 The first $N$ separated equations can be found by substituting $u_k$
 into (3.21)
  and using (3.37b), the last $N+1$
 separated equations are given by (3.37c) and (3.39a). They may be
 integrated to give
$$S(u_1,...,u_{2N+1})=\sum_{k=1}^{N}(2\int^{u_k}\sqrt {P(\la)}d\la
+2P_{N+k}u_{N+1+k})+P_{0} u_{N+1}, \tag 3.41$$
with $P(\la)$ given by (3.21).
Then the Jacobi inversion problem for the FDIHS (3.17) is
$$\sum_{k=1}^{N}\int^{u_k}\frac {1}{\sqrt {P(\la)}}d\la+u_{N+1}
=\gamma_{0},$$
$$\sum_{k=1}^{N}\int^{u_k}\frac {1}{(\la-\la_i)\sqrt {P(\la)}}d\la
=\gamma_{i}+2x,\quad i=1,...,N,$$
$$\sum_{k=1}^{N}\int^{u_k}\frac {P_{N+i}}{(\la-\la_i)^2\sqrt {
P(\la)}}d\la
+u_{N+1+i}=\gamma_{N+i},\quad i=1,...,N. \tag 3.42$$
The Jacobi inversion problem for the FDIHS (3.19) is
$$\sum_{k=1}^{N}\int^{u_k}\frac {1}{\sqrt {P(\la)}}d\la+u_{N+1}
=\gamma_{0}-P_0t_2,$$
$$\sum_{k=1}^{N}\int^{u_k}\frac {1}{(\la-\la_i)\sqrt {P(\la)}}d\la
=\gamma_{i}+2\la_it_2,\quad i=1,...,N,$$
$$\sum_{k=1}^{N}\int^{u_k}\frac {P_{N+i}}{(\la-\la_i)^2\sqrt {
P(\la)}}d\la
+u_{N+1+i}=\gamma_{N+i}+2P_{N+i}t_2,\quad i=1,...,N. \tag 3.43$$
Finally we have
\proclaim {Proposition 8}
The Jacobi inversion problem for the AKNS equation (3.6) is
$$\sum_{k=1}^{N}\int^{u_k}\frac {1}{\sqrt {P(\la)}}d\la+u_{N+1}
=\gamma_{0}-P_0t_2,$$
$$\sum_{k=1}^{N}\int^{u_k}\frac {1}{(\la-\la_i)\sqrt {P(\la)}}d\la
=\gamma_{i}+2(x+\la_it_2),\quad i=1,...,N,$$
$$\sum_{k=1}^{N}\int^{u_k}\frac {P_{N+i}}{(\la-\la_i)^2\sqrt {
P(\la)}}d\la
+u_{N+1+i}=\gamma_{N+i}+2P_{N+i}t_2,\quad i=1,...,N. \tag 3.44$$
\endproclaim\par
If $\phi_{1j}, \psi_{2j}, q$ defined by (3.40) can be solved
from (3.44) by using the Jacobi inversion technique, then
$\phi_{2j}, \psi_{1j}$ and $r$ can be obtained from the equations in
(3.17)
by an algebraic calculation, respectively. Finally $(q,r)$
provides the solution to the AKNS equations (3.6).\par
(3) The above procedure can be applied to all high-order binary
constrained
flows (3.8) and and whole AKNS hierarchy (3.4).\par

\ \par
\subhead {4. The separation of variables for
the Kaup-Newell equations }
\endsubhead\par

\subhead {4.1 Binary constrained flows of the Kaup-Newell hierarchy}
\endsubhead\par
\ \par
For the Kaup-Newell spectral problem [31]
$$\phi_x=U(u,\la)\phi,\quad
 U(u, \lambda)
=\left( \matrix -\la^2&q\la\\r\la&\la^2\endmatrix\right),\quad
\phi=\binom {\phi_{1}}{\phi_{2}},\quad u=\binom {q}{r},\tag 4.1$$
take
$$\phi_{t_n}=V^{(n)}(u,\la)\phi,\qquad V^{(n)}(u, \la)=\sum_{i=0}^{n-1}
\left( \matrix a_{2i}\la^{2n-2i}&b_{2i+1}\la^{2n-2i-1}
\\c_{2i+1}\la^{2n-2i-1}&-a_{2i}\la^{2n-2i}\endmatrix\right)\tag 4.2$$
where
$$a_{0}=1,\quad a_{2}=-\frac 12 qr,\quad b_{1}=-q,\quad c_1=-r,\quad
b_{3}
=\frac 12(q^2r+q_x),\quad c_3=\frac{1}{2}(qr^2-r_x),...,$$
and in general $a_{2k+1}=b_{2k}=c_{2k}=0$
$$\binom{c_{2k+1}}{ b_{2k+1}}=L \binom{c_{2k-1}}{ b_{2k-1}}, \qquad
a_{2k}=\frac 12\p^{-1}(qc_{2k-1,x}+rb_{2k-1,x}),\qquad
 k=1,2,\cdots,\tag 4.3$$
$$L=\frac 12\left( \matrix \p-r\p^{-1}q\p&-r\p^{-1}r\p\\-q\p^{-1}q\p
&-\p-q\p^{-1}r\p\endmatrix\right).$$\par
Then the compatibility condition of equations
(4.1) and (4.2) gives rise
to the Kaup-Newell hierarchy [31]
$$u_{t_n}={\binom {q}{r}}_{t_n}
=J\binom {c_{2n-1}}{ b_{2n-1}}
=J\frac {\delta H_{2n-2}}{\delta u},\qquad n=1,2,\hdots, \tag 4.4$$
where the Hamiltonian $H_n$ and the Hamiltonian operator $J$ are given
by
$$J=\left( \matrix 0&\p\\\p&0\endmatrix\right),\qquad
H_{2n}=\frac{4a_{2n+2}-rc_{2n+1}-q b_{2n+1}}{2n},\qquad
\binom {c_{2n+1}}{b_{2n+1}}
=\frac {\delta H_{2n}}{\delta u}.$$\par
For $n=2$ we have
$$\phi_{t_2}=V^{(2)}(u,\la)\phi,\qquad
V^{(2)}=\left( \matrix \la^4-\frac
12qr\la^2&-q\la^3+\frac{1}{2}(q^2r+q_x)
\la\\-r\la^3+\frac{1}{2}(qr^2-r_x)\la&
-\la^4+\frac 12qr\la^2\endmatrix\right)\tag 4.5$$
and the coupled derivative nonlinear Schr$\ddot{\text o}$dinger (CDNS)
equations obtained from the equation (4.4) for $n=2$ read
$$q_{t_2}=\frac 12q_{xx}+\frac 12(q^2r)_x, \qquad r_{t_2}=-\frac
12r_{xx}+
\frac 12(r^2q)_x. \tag 4.6$$\par
The adjoint Kaup-Newell spectral problem is the equation (2.7) with $U$
given by (4.1).
We have [26]
$$\frac {\delta\la}{\delta u}=\binom{\frac {\delta\la}{\delta q}}
{\frac {\delta\la}{\delta r}}=Tr[\left(\matrix \phi_1\psi_1
& \phi_1\psi_2\\ \phi_2\psi_1&\phi_2\psi_2
\endmatrix\right)\frac {\p U(u,\la)}{\p u}]=\binom {\la\psi_1\phi_2}
{\la\psi_2\phi_1}.\tag 4.7$$
\par
The binary $x$-constrained flows of the Kaup-Newell hierarchy (4.4)
are defined by
$$ \Phi_{1,x}=-\La^2\Phi_{1}+q\La\Phi_{2},\qquad\Phi_{2,x}=r\La\Phi_{1}
+\La^2\Phi_{2}
,\tag 4.8a$$
$$ \Psi_{1,x}=\La^2\Psi_{1}-r\La\Psi_{2},\qquad\Psi_{2,x}=-q\La\Psi_{1}
-\La^2\Psi_{2},\tag 4.8b$$
$$\frac {\delta H_{k_0}}{\delta u}-
\sum_{j=1}^{N}\frac {\delta \lambda_{j}}{\delta u}
=\binom {c_{2k_0+1}}{ b_{2k_0+1}}-\frac 12
\binom {<\La\Psi_1,\Phi_2>}{ <\La\Psi_2,\Phi_1>}=0.\tag 4.8c$$
For $k_0=0$, we have
$$\binom {c_{1}}{ b_{1}}=-\binom {r}{q}=\frac 12\binom {
<\La\Psi_1,\Phi_2>}{ <\La\Psi_2,\Phi_1>}.\tag 4.9$$
By substituting (4.9) into (4.8a) and (4.8b), the first binary
$x$-constrained flow becomes a
FDHS
$$ \Phi_{1x}=\frac {\p F_1}{\p \Psi_1},\quad \Phi_{2x}=\frac {\p F_1}
{\p \Psi_2},\quad
 \Psi_{1x}=-\frac {\p F_1}{\p \Phi_1},\quad
 \Psi_{2x}=-\frac {\p F_1}{\p \Phi_2},\tag 4.10$$
with the Hamiltonian
$$F_1=<\La^2\Psi_2, \Phi_2>-<\La^2\Psi_1, \Phi_1>
-\frac 12<\La\Psi_2, \Phi_1><\La\Psi_1, \Phi_2>.$$\par
Under the constraint (4.9) and the FDHS (4.10), the binary
$t_2$-constrained
flow obtained from  (4.5) and its adjoint equation for $N$ distinct
reral numbers $\la_j$ can also be written as a FDHS
$$ \Phi_{1,t_2}=\frac {\p F_2}{\p \Psi_1},\quad \Phi_{2,t_2}=\frac
{\p F_2}{\p \Psi_2},\quad
 \Psi_{1,t_2}=-\frac {\p F_2}{\p \Phi_1},\quad
 \Psi_{2,t_2}=-\frac {\p F_2}{\p \Phi_2},\tag 4.11$$
with the Hamiltonian
$$F_2=-<\La^4\Psi_2, \Phi_2>+<\La^4\Psi_1, \Phi_1>
+\frac 12<\La\Psi_2, \Phi_1><\La^3 \Psi_1, \Phi_2>$$
$$+\frac 12<\La^3 \Psi_2, \Phi_1><\La\Psi_1, \Phi_2>-\frac 1{32}
<\La \Psi_2, \Phi_1>^2<\La\Psi_1, \Phi_2>^2$$
$$+\frac 18(<\La^2\Psi_2, \Phi_2>-<\La^2\Psi_1, \Phi_1>)<\La\Psi_2,
\Phi_1>
<\La\Psi_1, \Phi_2>.$$\par
The Lax representation for the FDHSs (4.10) and (4.11) are presented
by (2.13) with the entries of the
 Lax matrix $M$ given by
$$A(\la)=1+\frac{1}{4}\sum_{j=1}^{N}\frac{\la_j^2(\psi_{1j}\phi_{1j}-
\psi_{2j}\phi_{2j})}
{\la^2-\la^2_{j}},\tag 4.12a$$
$$B(\la)=\frac 12\la\sum_{j=1}^{N}\frac{\la_j\psi_{2j}\phi_{1j}}{\la^2
-\la^2_{j}},\qquad
C(\la)=\frac 12\la\sum_{j=1}^{N}\frac{\la_j\psi_{1j}\phi_{2j}}{\la^2
-\la^2_{j}}. \tag 4.12b$$
A straightforward calculation yields
$$A^2(\la)+B(\la)C(\la)\equiv P(\la)=1+
\sum_{j=1}^{N}[\frac{P_{j}}{\la^2-\la^2_{j}}+\frac{\la_j^4P^2_{N+j}}
{(\la^2-\la^2_{j})^2}], \tag 4.13$$
where $P_j, j=1,...,2N,$ are $2N$ independent integrals of motion for
the
FDHSs (4.10) and (4.11)
$$P_j=-\frac 12\la_j^2(\psi_{2j}\phi_{2j}-\psi_{1j}\phi_{1j})+\frac 18
<\La\Psi_2, \Phi_1>\la_j\psi_{1j}\phi_{2j}+\frac 18<\La\Psi_1, \Phi_2>
\la_j\psi_{2j}\phi_{1j}$$
$$+\frac 18\sum_{k\neq j}\frac{1}{\la^2_j-\la^2_{k}}[\la^2_j\la^2_{k}
(\psi_{1j}\phi_{1j}-\psi_{2j}\phi_{2j})(\psi_{1k}\phi_{1k}
-\psi_{2k}\phi_{2k})
+2\la_j\la_{k}(\la^2_j+\la^2_{k})\psi_{1j}\phi_{2j}\psi_{2k}\phi_{1k}],$$

$$\qquad j=1,...,N\tag 4.14a$$
$$P_{N+j}=\frac 14(\psi_{1j}\phi_{1j}+\psi_{2j}\phi_{2j}),\qquad
j=1,...,N.\tag 4.14b$$
It is easy to varify that
$$F_1=-2\sum_{j=1}^{N}P_{j},\qquad
F_2=2\sum_{j=1}^{N}(\la_j^2P_{j}+\la_j^4P^2_{N+j})-\frac 12
(\sum_{j=1}^{N}P_{j})^2
, \tag 4.15a$$\par
$$<\Psi_2, \Phi_2>+<\Psi_1, \Phi_1>=4\sum_{j=1}^{N}P_{N+j}.\tag 4.15b$$
By inserting $\la=0$, (4.13) leads to
$$1+\frac 14(<\Psi_2, \Phi_2>-<\Psi_1, \Phi_1>)=\sqrt{P(0)}
=\sqrt{1+\sum_{j=1}^{N}[-P_{j}\la^{-2}_{j}+P^2_{N+j}]}.
\tag 4.16$$\par
With respect to the standard Poisson bracket it is found that
$$\{A(\la), A(\mu)\}=\{B(\la), B(\mu)\},\tag 4.17a$$
$$\{A(\la), B(\mu)\}=\frac {\mu}{2(\la^2-\mu^2)}
[\mu B(\mu)-\la B(\la)].
\tag 4.17b$$
Then $\{A^2(\la)+B(\la)C(\la),  A^2(\mu)+B(\mu)C(\mu)\}=0$  implies
that $P_j, j=1,...,2N,$ are in involution. The CDNS equations (4.6)
are factorized by the $x$-FDIHS (4.10) and the $t_2$-FDIHS (4.11),
namely,
if $(\Psi_1, \Psi_2, \Phi_1,\Phi_2)$
satisfies the FDIHSs (4.10) and (4.11) simultaneously,
then $(q, r)$ given by (4.9) solves the CDNS equations (4.6).
The factorization of the n-th Kaup-Newell ewuations (4.4) will be
presented in the end of section 4.2.\par
\ \par
\subhead {4.2 The separation of variables for the
Kaup-Newell equations}
\endsubhead\par
\ \par
Since the commutator relations (4.17) are quite different from (2.31)
and (3.23), we have to  modify a little bit of the method presented
in sections 2 and 3. Let us denote $\widetilde \la=\la^2,\widetilde
\la_j
=\la_j^2$. The entries of the Lax matrix $M$ given by (4.12) can be
rewritten as
$$A(\widetilde\la)=1+\frac 14(<\Psi_2, \Phi_2>-<\Psi_1, \Phi_1>)+
\frac 12 \widetilde\la \overline
A(\widetilde\la),\quad B(\widetilde\la)
=\frac 12\sqrt{\widetilde\la}\overline
B(\widetilde\la),\tag 4.18a$$
where
$$\overline A(\widetilde\la)=\frac{1}{2}\sum_{j=1}^{N}\frac{\psi_{1j}
\phi_{1j}-\psi_{2j}\phi_{2j}}{\widetilde\la-\widetilde\la_{j}},\quad
\overline B(\widetilde\la)=\sum_{j=1}^{N}\frac{\sqrt{\widetilde\la_j}
\psi_{2j}\phi_{1j}}{\widetilde\la-\widetilde\la_{j}}.\tag 4.18b$$
It is easy to see that
$$\{\overline A(\widetilde\la), \overline A(\widetilde\mu)\}
=\{\overline B(\widetilde\la), \overline B(\widetilde\mu)\}=0,\tag
4.19a$$
$$\{\overline A(\widetilde\la), \overline B(\widetilde\mu)\}
=\frac 1{\widetilde\la-\widetilde\mu}[\overline B(\widetilde\mu)
-\overline B(\widetilde\la)].\tag 4.19b$$
It follows from (4.16) and (4.18a) that
$$A(\widetilde\la)
=\sqrt{1+\sum_{j=1}^{N}[-P_{j}\widetilde\la^{-1}_{j}+P^2_{N+j}]}
+\frac 12 \widetilde\la \overline A(\widetilde\la).\tag 4.19c$$
The commutator relations (4.19) and the generating function of
integrals of motion (4.13) enable us to introduce
$u_1,...,u_N$ in the following way:
$$\overline B(\widetilde\la)=\sum_{j=1}^{N}\frac{\sqrt{\widetilde\la_j}
\psi_{2j}\phi_{1j}}{\widetilde\la-\widetilde\la_{j}}
=e^{u_N}\frac {R(\widetilde\la)}{K(\widetilde\la)},\tag 4.20a$$
with
$$R(\widetilde\la)=\prod_{k=1}^{N-1}(\widetilde\la-u_{k}),\qquad
K(\widetilde\la)=\prod_{k=1}^{N}(\widetilde\la-\widetilde\la_{k}),$$
and $v_1,...,v_N$ by
$\overline A(\widetilde \la)$:
$$v_{k}=\overline A(u_k),k=1,...,N-1,\tag 4.20b$$
$$v_{N}=-<\Psi_{2}, \Phi_{2}>.\tag 4.20c$$
The eq. (4.20a) yields
$$u_N=\text {ln}<\La\Psi_{2}, \Phi_{1}>.\tag 4.20d$$
Similarly we define
$$v_{N+j}=2P_{N+j}, \quad j=1,...,N,\tag 4.21a$$
$$u_{N+j}=\text {ln}\frac {\phi_{1j}}{\psi_{2j}},\quad j=1,...,N.
\tag 4.21b$$
Then we have
\proclaim {Proposition 9}
Assume that  $\la_j, \phi_{ij}, \psi_{ij} \in\text {\bf R}, i=1,2,
j=1,...,N$.
Introduce the separated variables
$u_{1},...,u_{2N}$ and  $v_{1},...,v_{2N}$ by (4.20) and (4.21).
If $u_1,...,u_{N-1},$ are single zeros of $\overline B(\la)$, then
$v_1,...,v_{2N}$ and $u_1,...,u_{2N}$ are canonically conjugated, i.e.,
they  satisfy (1.1).\par
\endproclaim\par
It follows from (4.9), (4.20a), (4.20d)  and (4.21b) that
$$q=-\frac 12e^{u_{N}},\tag 4.22a$$\par
$$\phi_{1j}=\sqrt{\frac{e^{u_{N}+u_{N+j}}
R(\la_j^2)}{\la_jK'(\la_{j}^2)}},
\qquad
\psi_{2j}=\sqrt{\frac{e^{u_{N}-u_{N+j}}
R(\la_j^2)}{\la_jK'(\la_{j}^2)}},
,\qquad j=1,...,N.\tag 4.22b$$\par
By substituting $u_k$ into (4.13) and using (4.16) and (4.19c),
one gets
the first $N-1$  separated equations
$$v_k=\overline A(u_k)=\frac 2{u_k}[\sqrt {\widetilde P(u_k)}
-\sqrt{P(0)}],
\quad k=1,...,N-1, \tag 4.23a$$
where $P(0)$ are given by (4.16) and
$$\widetilde P(\widetilde\la)=
1+\sum_{j=1}^{N}[\frac{P_{j}}{\widetilde\la-\la^2_{j}}+\frac
{\la_j^4P^2_{N+j}}{(\widetilde\la-\la^2_{j})^2}].$$
It follows from (4.15b), (4.16) and (4.20c) that
$$v_N=2-2\sqrt{P(0)}-2\sum_{i=1}^{N}P_{N+i}.\tag 4.23b$$
The separated equations (4.23) and (4.21a) may be
integrated to give the
generating function of the canonical transformation
$$S(u_1,...,u_{2N})
=\sum_{k=1}^{N-1}[\int^{u_k}\frac 2{\widetilde\la}\sqrt
{\widetilde P(\widetilde\la)}d\widetilde\la-2\sqrt{P(0)}ln|u_k|]$$
$$+(2-2\sqrt{P(0)}-2\sum_{i=1}^{N}P_{N+i})u_N+2
\sum_{i=1}^{N}P_{N+i}
u_{N+i}.
 \tag 4.24$$\par
The Jacobi inversion problem for the FDIHS (4.10) is
$$\sum_{k=1}^{N-1}[\int^{u_k}\frac {1}{\widetilde\la(\widetilde\la
-\la_i^2)\sqrt {\widetilde P(\widetilde\la)}}d\widetilde\la+\frac 1
{\la_i^2\sqrt{P(0)}}ln|u_k|]
+\frac 1{\la_i^2\sqrt{P(0)}}u_N=\gamma_i-2x,$$
$$\sum_{k=1}^{N-1}[\int^{u_k}\frac {\la_i^4P_{N+i}}{\widetilde
\la(\widetilde\la-\la_i^2)^2\sqrt {\widetilde P(\widetilde\la)}}
d\widetilde\la-\frac {P_{N+i}}{\sqrt{P(0)}}ln|u_k|]$$
$$-(\frac {P_{N+i}}{\sqrt{P(0)}}+1)u_{N}+u_{N+i}=\gamma_{N+i},
 \quad i=1,...,N. \tag 4.25$$\par
The Jacobi inversion problem for the FDIHS (4.11) is
$$\sum_{k=1}^{N-1}[\int^{u_k}\frac {1}{\widetilde\la(\widetilde\la
-\la_i^2)\sqrt {\widetilde P(\widetilde\la)}}d\widetilde\la
+\frac 1{\la_i^2\sqrt{P(0)}}ln|u_k|]
+\frac 1{\la_i^2\sqrt{P(0)}}u_N$$
$$=\bar\gamma_i+(2\la_i^2-\sum_{k=1}^{N}P_k)t_2, $$
$$\sum_{k=1}^{N-1}[\int^{u_k}\frac {\la_i^4P_{N+i}}{\widetilde
\la(\widetilde\la-\la_i^2)^2\sqrt {\widetilde P(\widetilde\la)}}
d\widetilde\la-\frac {P_{N+i}}{\sqrt{P(0)}}ln|u_k|]$$
$$-(\frac {P_{N+i}}{\sqrt{P(0)}}+1)u_{N}+u_{N+i}
=\bar\gamma_{N+i}+2\la_i^4P_{N+i}t_2, \quad \qquad i=1,...,N. \tag
4.26$$
\par
Finally, since the CDNS equations (4.6) are factorized by the FDIHS
(4.10)
and (4.11),  combining the equation (4.25) with the equation (4.26)
gives rise to the
Jacobi inversion problem for the CDNS equations (4.6)
$$\sum_{k=1}^{N-1}[\int^{u_k}\frac {1}{\widetilde\la(\widetilde\la
-\la_i^2)\sqrt {\widetilde P(\widetilde\la)}}d\widetilde\la+\frac
1{\la_i^2\sqrt{P(0)}}ln|u_k|]
+\frac 1{\la_i^2\sqrt{P(0)}}u_N$$
$$=\gamma_i-2x+(2\la_i^2-\sum_{k=1}^{N}P_k)t_2,
\qquad \quad i=1,...,N, \tag 4.27a$$
$$\sum_{k=1}^{N-1}[\int^{u_k}\frac {\la_i^4P_{N+i}}{\widetilde
\la(\widetilde\la-\la_i^2)^2\sqrt {\widetilde P(\widetilde\la)}}
d\widetilde\la-\frac {P_{N+i}}{\sqrt{P(0)}}ln|u_k|]$$
$$-(\frac {P_{N+i}}{\sqrt{P(0)}}+1)u_{N}+u_{N+i}=\gamma_{N+i}
+2\la_i^4P_{N+i}t_2, \quad \quad i=1,...,N. \tag 4.27b$$
If $\phi_{1j}, \psi_{2j}, q$ defined by (4.22) can be solved from (4.36)

by using the Jacobi inversion technique, then $\phi_{2j}, \psi_{1j}$ can

be obtained from the first equation  and the last equation in (4.10),
respectively. Finally $q$ and $r=-<\La\Psi_1,\Phi_2>$ provides the
solution to the CDNS equations (4.6).\par
In general, the above procedure can be applied to the whole Kaup-Newell
hierarchy (4.4).
Set
$$A^2(\la)+B(\la)C(\la)=
\sum_{k=0}^{\infty}\widetilde F_k\la^{-2k}, \tag 4.28a$$
where $\widetilde F_k, k=1,2,...,$ are also integrals of motion
for both the $x$-FDHSs (4.10) and the $t_n$-binary constrained flows
(2.16).
Comparing (4.28a) with (4.13), one gets
$$\widetilde F_0=1, \quad \widetilde F_k=\sum_{j=1}^{N}[\la_j^{2k-2}P_j
+(k-1)\la_j^{2k}P_{N+j}^2],\quad k=1,2,.... \tag 4.28b$$
By employing the method in [28,29], the $t_n$-FDIHS obtained from
the $t_n$-constrained flow  is  of the form
$$ \Phi_{1,t_n}=\frac {\p F_{n}}{\p \Psi_1},\quad \Phi_{2,t_n}
=\frac {\p F_{n}}{\p \Psi_2},\quad
 \Psi_{1,t_n}=-\frac {\p F_{n}}{\p \Phi_1},\quad
 \Psi_{2,t_n}=-\frac {\p F_{n}}{\p \Phi_2},\tag 4.29a$$
with the Hamiltonian
$$F_{n}=2\sum_{m=0}^{n-1}(-\frac 12)^{m}\frac{\alpha_m}{m+1}
\sum_{l_1+...+l_{m+1}=n}\widetilde F_{l_1}...\widetilde F_{l_{m+1}},
\tag 4.29b$$
where $l_1\geq 1,...,l_{m+1}\geq 1,$ and  $\alpha_m$ are given by
(2.22).
Since the n-th Kaup-Newell equations (4.4) is factorized by the
$x$-FDIHS (4.10)
and the $t_n$-FDIHS (4.29). We have the
following proposition.
\proclaim {Proposition 10} The Jacobi inversion problem for the
n-th Kaup-Newell equations (4.4) is given by
$$\sum_{k=1}^{N-1}[\int^{u_k}\frac {1}{\widetilde\la(\widetilde\la
-\la_i^2)\sqrt {\widetilde P(\widetilde\la)}}d\widetilde\la
+\frac 1{\la_i^2\sqrt{P(0)}}ln|u_k|]
+\frac 1{\la_i^2\sqrt{P(0)}}u_N
=\gamma_i-2x$$
$$+2t_n\sum_{m=0}^{n-1}(-\frac 12)^m\alpha_m\sum_{l_1+...+l_{m+1}
=n}\la_i^{2l_{m+1}-2}\widetilde F_{l_1}...\widetilde F_{l_{m}},
\quad i=1,...,N, \tag 4.30a$$
$$\sum_{k=1}^{N-1}[\int^{u_k}\frac {\la_i^4P_{N+i}}{\widetilde
\la(\widetilde\la-\la_i^2)^2\sqrt {\widetilde P(\widetilde\la)}}
d\widetilde\la-\frac {P_{N+i}}{\sqrt{P(0)}}ln|u_k|]
-(\frac {P_{N+i}}{\sqrt{P(0)}}+1)u_{N}+u_{N+i}=\gamma_{N+i}$$
$$+2t_n\sum_{m=0}^{n-1}(-\frac 12)^m\alpha_m
\sum_{l_1+...+l_{m+1}=n}(l_{m+1}-1)\la_i^{2l_{m+1}}P_{N+i}
\widetilde F_{l_1}...\widetilde F_{l_{m}},
\quad i=1,...,N, \tag 4.30b$$
where $l_1\geq 1,...,l_{m+1}\geq 1,$ and $\widetilde F_{l_1},...
\widetilde F_{l_{m}},$ are given by (4.28b).
\endproclaim\par
For example,
the third equations in the Kaup-Newell hierarchy with $n=3$ are of the
form
$$q_{t_3}=-\frac 14 q_{xxx}-\frac 38(q^3r^2+2qrq_x)_x,\quad
r_{t_3}=-\frac 14 r_{xxx}-\frac 38(r^3q^2-2qrr_x)_x.\tag 4.31$$
The Kaup-Newell equations (4.31) can be factorized by the $x$-FDIHS
(4.10)
and $t_3$-FDIHS with the Hamiltonian $F_3$ defined by
$$F_3=\sum_{j=1}^{N}(2\la_j^4P_{j}+4\la_j^6P^2_{N+j})
-[\sum_{j=1}^{N}(\la_j^2P_{j}+\la_j^4P^2_{N+j})]\sum_{j=1}^{N}P_{j}
+\frac 14(\sum_{j=1}^{N}P_{j})^3. \tag 4.32$$
The Jacobi inversion problem for the equations (4.31) is given by
$$\sum_{k=1}^{N-1}[\int^{u_k}\frac {1}{\widetilde\la(\widetilde\la
-\la_i^2)\sqrt {\widetilde P(\widetilde\la)}}d\widetilde\la
+\frac 1{\la_i^2\sqrt{P(0)}}ln|u_k|]
+\frac 1{\la_i^2\sqrt{P(0)}}u_N$$
$$=\gamma_i-2x+[2\la_i^4-\sum_{j=1}^{N}(\la_j^2P_{j}+\la_i^2P_{j}
+\la_j^4P^2_{N+j})+\frac 34(\sum_{j=1}^{N}P_j)^2]t_3,
\qquad \quad i=1,...,N, $$
$$\sum_{k=1}^{N-1}[\int^{u_k}\frac {\la_i^4P_{N+i}}{\widetilde
\la(\widetilde\la
-\la_i^2)^2\sqrt {\widetilde P(\widetilde\la)}}d\widetilde\la-
\frac {P_{N+i}}{\sqrt{P(0)}}ln|u_k|]
-(\frac {P_{N+i}}{\sqrt{P(0)}}+1)u_{N}$$
$$+u_{N+i}=\gamma_{N+i}+[4\la_i^6P_{N+i}
-\la_i^4P_{N+j}\sum_{j=1}^{N}P_{j}]t_3, \quad \quad i=1,...,N.$$\par
In general, the method can be applied to all high-order binary
constrained
flows (4.8) and whole KN hierarchy (4.4) in the exactly same way.\par

\subhead {4. Concluding remarks}\endsubhead\par

For high-order binary constrained flows, the method in [1-6] allows us
to directly introduce  $N+k_0$ pairs
of canonical separated variables and $N+k_0$ separated
equations via the Lax matrices and generating function of integrals of
motion.
In this paper we propose a new method
for determining  additional $N$ pairs of canonical separated variables
and $N$ additional
separated equations for high-order binary constrained flows by directly
using $N$
additional integrals of motion. This method is completely different
from that proposed in [23,24] and can be applied to all high-order
binary
constrained flows and  other soliton hierarchies admitting $2\times 2$
Lax
pairs.\par

\subhead {Acknowledgments}\endsubhead\par
The author is grateful to Professor V.B. Kuznetsov for useful
discussions.
This work was supported by  the Chinese Basic Research Project
``Nonlinear
Science''.

\par
\  \par

\subhead {References}\endsubhead \par

\item  {1.} Sklyanin E.K. 1989 J. Soviet. Math. 47, 2473.\par
\item  {2.} Kuznetsov V.B. 1992 J. Math. Phys. 33, 3240.\par
\item  {3.} Sklyanin E.K. 1995 Prog. Theor. Phys. Suppl. 118, 35\par
\item  {4.} Eilbeck J.C., Enol'skii V.Z., Kuznetsov V.B. and Tsiganov
A.V. 1994 J. Phys. A: Math. Gen. 27, 567.\par
\item  {5.} Kalnins E.G., Kuznetsov V.B. and Willard Miller Jr 1994
J. Math. Phys. 35, 1710.\par
\item  {6.} Harnad J. and Winternitz P. 1995 Commun. Math. Phys. 172,
263.\par
\item  {7.} Dubrovin B.A. 1981 Russian Math. Survey 36, 11. \par
\item {8.} Krichever I.M. and Novikov S.P. 1980 Russian Math. Surveys
32,
53.\par
\item {9.} Adler M. and van Moerbeke P. 1980 Adv. Math. 38, 267.\par
\item  {10.} Kulish P.P., Rauch-Wojciechowski S. and Tsiganov A.V.
1996 J. Math. Phys. 37, 3463.\par
\item  {11.} Zeng Yunbo 1996 Phys. Lett. A 216, 26.\par
\item {12.} Zeng Yunbo 1997 J. Math. Phys. 38, 321.\par
\item{13.} Zeng Yunbo 1997 J. Phys. A: Math. Gen. 30, 3719.\par
\item{14.} Zeng Yunbo 1997 J. Phys. Soc. Jpn. 66, 2277.\par
\item{15.} Ragnisco O. and  Rauch-Wojciechowski S. 1992
Inverse Problems 8, 245.\par
\item{16.} Zeng Yunbo 1992 Chinese Science Bulletin 37, 1937.\par
\item{17.} Ma W.X. and Strampp W. 1994 Phys. Lett. A 185, 277.\par
\item{18.} Ma W.X. 1995 J. Phys. Soc. Jpn. 64, 1085.\par
\item{19.} Ma W.X., Fuchssteiner B. and Oevel W. 1996 Physica A 233,
331.\par
\item{20.} Ma W.X., Ding Q., Zhang W.G. and Lu B.Q. 1996
IL Nuovo Cimento B 111, 1135.\par
\item{21.} Li Yishen and Ma W.X. Binary nonlinearization of AKNS
spectral
problem under higher-order symmetry constraints,
to appear in Chaos, Solitons and Fractals.\par
\item {22.} Ma W.X. and Fuchssteiner B. 1996 in: Nonlinear
Physics, ed. E. Alfinito, M. Boiti, L. Martina and F. Pempinelli,
(Singapore: World Scientific) p217.\par
\item {23.} Zeng Yunbo and Ma W.X. The construction of canonical
separated variables for binary constrained AKNS flow, to appear
in Physica A.\par
\item {24.} Zeng Yunbo and Ma W.X. Seaparation of variables for
soliton equations via their binary constrained flows, to appear in J.
Math.
Phys.            \par
\item {25.} Newell A.C. 1985 Solitons in mathematics and physics
(Philadelphia: SIAM).\par
\item {26.} Tu Guizhang, Gradients and coadjoint representation
equations
of isospectral problem, preprint.\par
\item{27.} Zeng Yunbo and Li Yishen 1993 J. Phys. A: Math. Gen. 26,
L273.\par
\item{28.} Zeng Yunbo 1994 Physica D 73, 171.\par
\item{29.} Zeng Yunbo 1991 Phys. Lett. A 160, 541.\par
\item{30.} Ablowitz M. and Segur H. 1981
Solitons and the inverse scattering transform (Philadelphia: SIAM).\par
\item {31.} Kaup D.J. and Newell A.C.
1978 J. Math. Phys. 19, 798.\par

\bye
\bye